\documentclass[runningheads]{llncs}
\usepackage{graphicx}

\usepackage{float}
\usepackage{orcidlink}

\begin{document}
\title{AI-Enhanced Ethical Hacking: A Linux-Focused Experiment} 


%
\titlerunning{AI-Enhanced Ethical Hacking}
%

\author{Haitham S. Al-Sinani\orcidlink{0009-0005-0453-3335} \and Chris J. Mitchell\orcidlink{0000-0002-6118-0055}}

\authorrunning{H. Al-Sinani \& C. Mitchell}
%
\institute{Department of Cybersecurity and Quality Control, Diwan of Royal Court, Muscat, Oman. \email{hsssinani@diwan.gov.om} \and
 Department of Information Security, Royal Holloway, University of London, Egham, Surrey. TW20 0EX, UK. \email{C.Mitchell@rhul.ac.uk}\\
}
\maketitle              
\begin{abstract}

This technical report investigates the integration of generative AI (GenAI), specifically ChatGPT, into the practice of ethical hacking through a comprehensive experimental study and conceptual analysis. Conducted in a controlled virtual environment, the study evaluates GenAI's effectiveness across the key stages of penetration testing on  \textbf{Linux-based}
 target machines operating within a virtual local area network (LAN), including reconnaissance, scanning and enumeration, gaining access, maintaining access, and covering tracks.  The findings confirm that GenAI can significantly enhance and streamline the ethical hacking process while underscoring the importance of balanced human-AI collaboration rather than the complete replacement of human input. The report also critically examines potential risks such as misuse, data biases, hallucination, and over-reliance on AI. This research contributes to the ongoing discussion on the ethical use of AI in cybersecurity and highlights the need for continued innovation to strengthen security defences.


\end{abstract}
\section{Introduction}
\label{Introduction}
Ethical hacking~\cite{NIST800-115} is a crucial aspect of modern cybersecurity, yet it remains a highly time-consuming and resource-intensive endeavour. It requires not only advanced expertise but also continuous knowledge updates to stay ahead of rapidly evolving threats. Traditional ethical hacking approaches demand significant human involvement at each phase, from reconnaissance to vulnerability scanning and  exploitation, which increases both the time and overall costs involved.

Additionally, the process relies heavily on skilled professionals to effectively identify and exploit vulnerabilities, making it challenging to keep up with the growing scale and sophistication of attacks. These efforts are further constrained by the limited capacity of human operators to manage complex or large-scale environments without substantial investments in training and resources.

The integration of AI technologies, particularly GenAI, offers a promising solution to the challenges faced in ethical hacking by automating and enhancing various stages of the process. Tools like ChatGPT\footnote{\url{https://openai.com/blog/chatgpt}} \cite{brown2020language} allow ethical hackers to streamline repetitive tasks, make faster decisions, and reduce the extensive human input typically required. This not only addresses the time and capacity limitations faced by operators but also lowers the implementation costs. GenAI’s ability to analyse data, provide real-time insights, and optimise workflows leads to more efficient, cost-effective security assessments.

%
%

This report presents a comprehensive experimental study evaluating the practical use of GenAI in a controlled Linux-based virtual environment. By simulating key stages of ethical hacking, such as reconnaissance, scanning, gaining \& maintaining access, and covering tracks, this study demonstrates how GenAI can enhance these processes and bolster cybersecurity defences. The findings and observations documented here contribute to the ongoing discussion about AI-human collaboration in cybersecurity, emphasising the potential of GenAI to improve efficiency and reduce costs while maintaining the need for expert oversight.

While previous research has explored the broader role of GenAI in cybersecurity, this report specifically examines its application in Linux-based environments, which are frequently targeted in both penetration testing and real-world attacks.
This work builds on our previously published research, in which we proposed a conceptual model
leveraging the capabilities of GenAI to support ethical hackers across the five stages of ethical
hacking~\cite{STM24_UnleashingAIinEthicalHacking}. It also expands on a proof-of-concept
implementation, used to conduct an initial experimental study on the integration of AI into ethical
hacking on target Windows  VMs~\cite{TechReportUnAIInEH_HC_2024}.

The remainder of this document is organised as follows. Section~\ref{Generative AI and ChatGPT}
explores  GenAI and ChatGPT.  Section~\ref{Laboratory Setup} presents the laboratory setup, and
section~\ref{Methodology} outlines our methodology. Section~\ref{Execution} details the execution
of our experiment. Section~\ref{DiscussionAndAnalysis} discusses the potential benefits and risks.
Section~\ref{Related work} reviews related work, and, section~\ref{Conclusions and future work}
summarises our conclusions and outlines plans for future work. Finally, appendix~\ref{Appendix}
lists  all the figures referenced in this technical report.

\section{Generative AI and ChatGPT}
\label{Generative AI and ChatGPT} The advent of GenAI, with models like
ChatGPT\footnote{\url{https://openai.com/blog/chatgpt}} \cite{brown2020language} prominent,
represents a major shift in the AI landscape. These systems, moving beyond the traditional AI focus
on pattern recognition and decision-making, excel in content creation, including text, images, and
code. The ability to learn from extensive datasets and produce outputs that mimic human creativity
is a major advance.

Central to this revolution is the GPT (Generative Pre-trained Transformer) architecture, the basis
of models like ChatGPT\@. Developed by OpenAI, GPT models are built on deep learning techniques
using \textit{transformer} models, designed specifically for handling sequential data. These models
undergo pre-training, where they learn from a wide array of various resources, including Internet
texts, followed by fine-tuning for specific tasks. This process enables models to grasp not just
the structure of language but also its context, essential for generating human-like text.

Each iteration of ChatGPT has demonstrated enhanced contextual understanding and output relevance.
Its primary function lies in interpreting user prompts and generating coherent, contextually
appropriate responses. This versatility extends from conducting conversations to performing complex
tasks, including coding, content creation, and, as we propose in this report, ethical hacking. The
GPT model family, including ChatGPT, owes much of its success to the transformer model, introduced
by Vaswani et al.\ in 2017 \cite{vaswani2017attention}. This architecture revolutionises sequence
processing through attention mechanisms, enabling the model to focus on different parts of the
input based on its relevance to the task.

The latest iteration, GPT-4o\footnote{\url{https://openai.com/blog/chatgpt}}, provides significant
advances in speed, multimodal capabilities, and overall intelligence. GPT-4o, now available to a
broader user base, including free-tier users, improves upon the GPT-4 model by offering enhanced
performance in understanding and generating text, as well as new capabilities in processing voice
and images. These improvements position GPT-4o as a powerful tool not only in natural language
processing but also in applications such as real-time communication and data analysis, making it a
key asset in modern cybersecurity practices.

In exploring the intersection of AI and cybersecurity, understanding ChatGPT's foundational aspects
is vital. Its generative nature, contextual sensitivity, and adaptive learning capacity can lead to
innovative approaches in cybersecurity practices. Our focus will be on how these qualities of
ChatGPT can be used to support ethical hacking, exploring the technical, ethical, and practical
implications.

\section{Laboratory Setup}
\label{Laboratory Setup}

\subsection{Physical Host and Virtual Environment Configuration}
\label{Physical Host} \label{Virtual Environment Configuration}

\begin{figure}
\includegraphics[width=\textwidth]{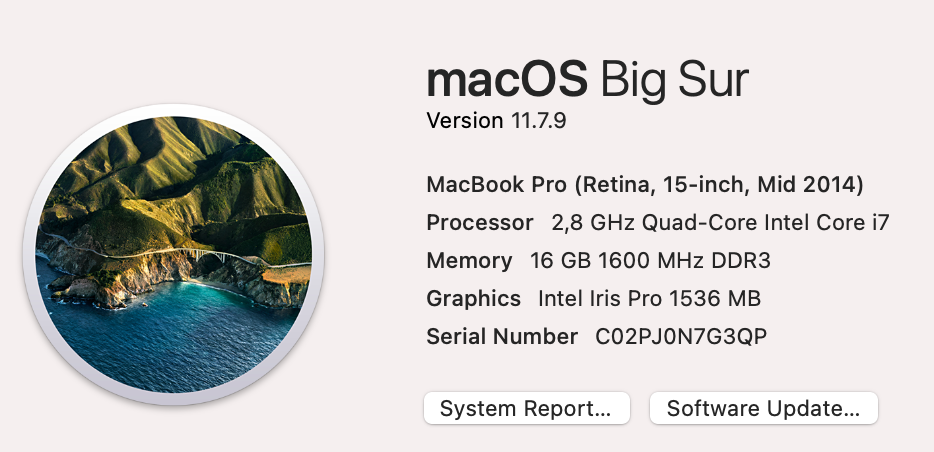}
\caption{MacBook: the physical host}
\label{macbook}
\end{figure}

\begin{figure}
\includegraphics[width=\textwidth]{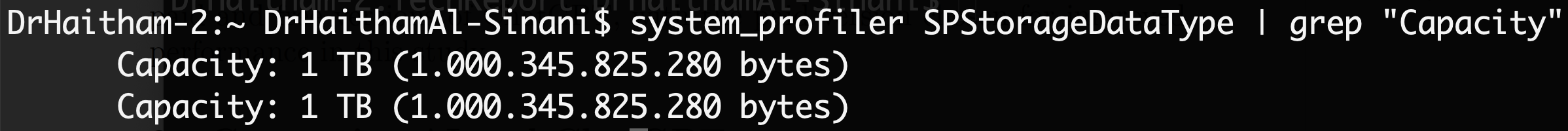}
\caption{MacBook size}
\label{macbook_size}
\end{figure}

The experiments used a MacBook Pro with 16 GB RAM, a 2.8 GHz Quad-Core Intel Core i7 processor, and
1 TB of storage, providing sufficient  computational capabilities for virtualisation (see
Figs.~\ref{macbook} and~\ref{macbook_size}).

\begin{figure}
\includegraphics[width=\textwidth]{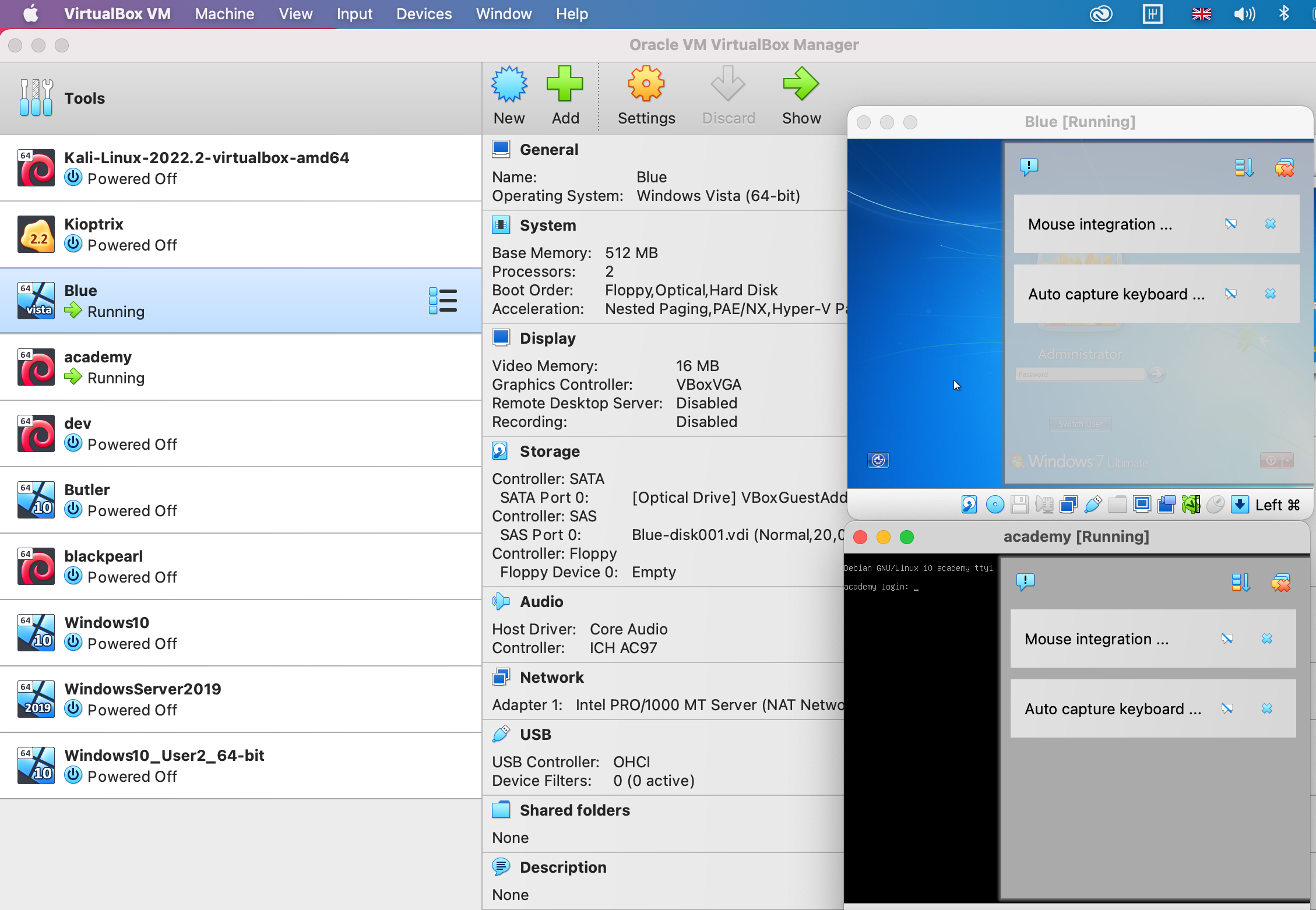}
\caption{VirtualBox \& VMs}
\label{VirtualBox_VMs}
\end{figure}

Virtualisation of the network was achieved using VirtualBox 7 (see Fig.~\ref{VirtualBox_VMs}), a
reliable tool for creating and managing virtual machine environments. The virtual setup included
the following  VMs.

\begin{enumerate}
 \item \textbf{Kali Linux VM:} this machine functioned as the primary attack platform for conducting the penetration tests. It is equipped with the necessary tools and applications for ethical hacking.
  \item \textbf{Windows VM:} this  machine, running a 64-bit version of Windows Vista with a memory allocation of 512 MB, was the principal target for penetration testing within a previously conducted  experiment~\cite{TechReportUnAIInEH_HC_2024}.
  \item \textbf{Linux VM:} this  machine, operating on a 64-bit Linux Debian system and allocated 512 MB of memory, is the primary focus of this report.
  \end{enumerate}

The network configuration was established in a local NAT (Network Address Translation)  
setup, allowing for seamless communication between the VMs and simulating a realistic network
environment suitable for penetration testing.

\subsection{Generative AI Tool}
\label{Generative AI Tool} The experiment leveraged
ChatGPT-4\footnote{\url{https://openai.com/index/hello-gpt-4o/}} (a paid version)    for its
advanced AI capabilities and efficient response time. The selection of ChatGPT-4 was also 
based on its prominent status as a leading GenAI tool, offering cutting-edge technology to enhance
the ethical hacking process. Of course, other GenAI tools are also available, e.g.\ Google's
Bard\footnote{\url{https://bard.google.com/}}  and GitHub's
Co-Pilot\footnote{\url{https://github.com/features/copilot/}},  which could potentially be used in
similar contexts. The methodologies and processes described are applicable to both the paid and
free versions of ChatGPT, with the paid version chosen for improved performance in this study.

\section{Methodology}
\label{Methodology} The experiment followed the structured phases of ethical hacking listed below,
with ChatGPT's guidance integrated at each step.

\begin{enumerate}
\item \textbf{Reconnaissance:} ChatGPT was used to gather and analyse information about the
    target VMs, including scanning to discover live machines.
\item \textbf{Scanning and Enumeration:} Network and vulnerability scanning were conducted using tools such as
    nmap, with ChatGPT helping to interpret the scan results and identify potential
    vulnerabilities.
\item \textbf{Gaining Access (Linux VM):}  This phase focused on exploiting identified vulnerabilities
    using the Metasploit framework. ChatGPT assisted in selecting and configuring the
    appropriate exploit.
\item \textbf{Maintaining \& Elevating Access:} ChatGPT suggested methods for maintaining
    access, such as creating backdoor accounts and escalating privileges within the compromised
    system.
\item \textbf{Covering Tracks \& Documentation:}  In the post-exploitation phase, ChatGPT
    advised on strategies to effectively erase traces of the penetration test, thereby reducing
    the likelihood of detection by system administrators. This included log manipulation and
    account removal. Additionally, ChatGPT assisted in documenting the ethical hacking process,
    ensuring comprehensive reporting of methodologies, findings, and recommendations for
    enhancing system security.
\end{enumerate}

\begin{figure}
\includegraphics[width=\textwidth]{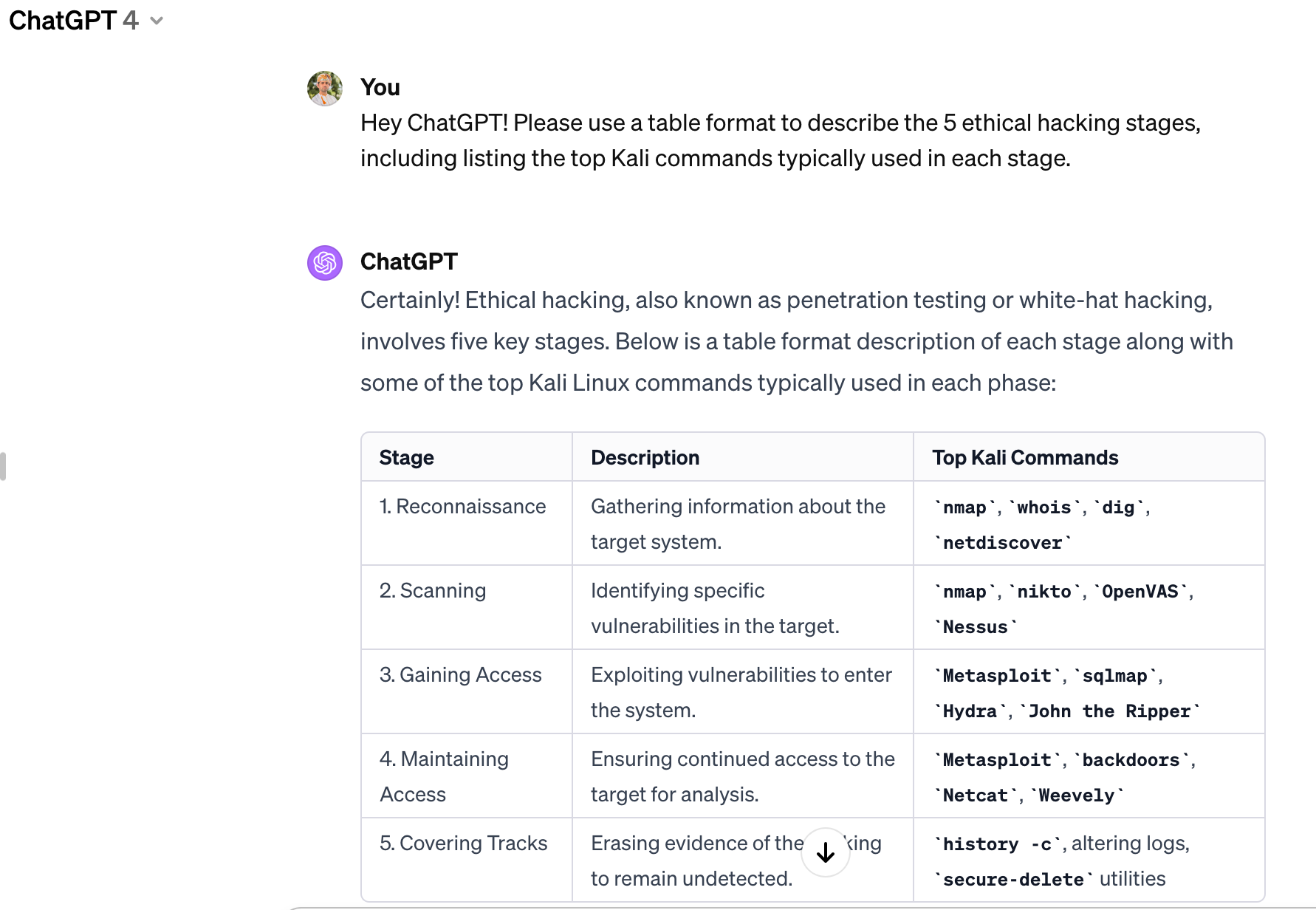}
\caption{ The five ethical hacking stages}
\label{tableEthicalHackingStages}
\end{figure}

We initiated the experiment by asking ChatGPT to provide a concise explanation of the five ethical
hacking stages, along with a list of commonly used Kali commands for each stage. ChatGPT provided
an informative response, as illustrated in Fig.~\ref{tableEthicalHackingStages}.

\section{Execution}
\label{Execution}

\label{Objective}

We now summarise the experimental procedure for each stage.


\subsection{Reconnaissance}
\label{Reconnaissance}

There are two main types of reconnaissance (recon).
\begin{enumerate}
\item \textbf{Passive Recon:} This entails passive observation without active engagement.
\item \textbf{Active Recon:} Active recon involves engaging with the target to prompt responses
    for observation.
\end{enumerate}

The emphasis here is on active reconnaissance; we followed the steps listed below.

\subsubsection{Notes on VM IP Address.} \label{VMIPAddress} First, observe that in our
VirtualBox-driven, NAT-based VM environment, the DHCP server is configured by default to
dynamically assign IP addresses to the VMs. DHCP typically allocates IP addresses sequentially
within the specified range. For instance, if the range is 192.168.1.0/24 (see
Fig.~\ref{NATspecifiedRange} below), the first IP address assigned would likely be 192.168.1.1
(often reserved for the default gateway), followed by 192.168.1.2, and so on. However, IP addresses
may change between device sessions. The availability of a specific IP address depends on several
factors, including the DHCP lease time and the currently active VMs. To maintain consistency in the
experiment, and because the originally assigned dynamic IP address for the Linux VM mentioned
earlier (192.168.1.7) had changed, we opted to assign a static IP address, reverting it to
192.168.1.7, as shown in Fig.~\ref{StaticIPaddressAssignment}  below. While this approach does not
scale well for large, enterprise-level networks, it is practical in our controlled research
environment.

\begin{figure}
\includegraphics[width=\textwidth]{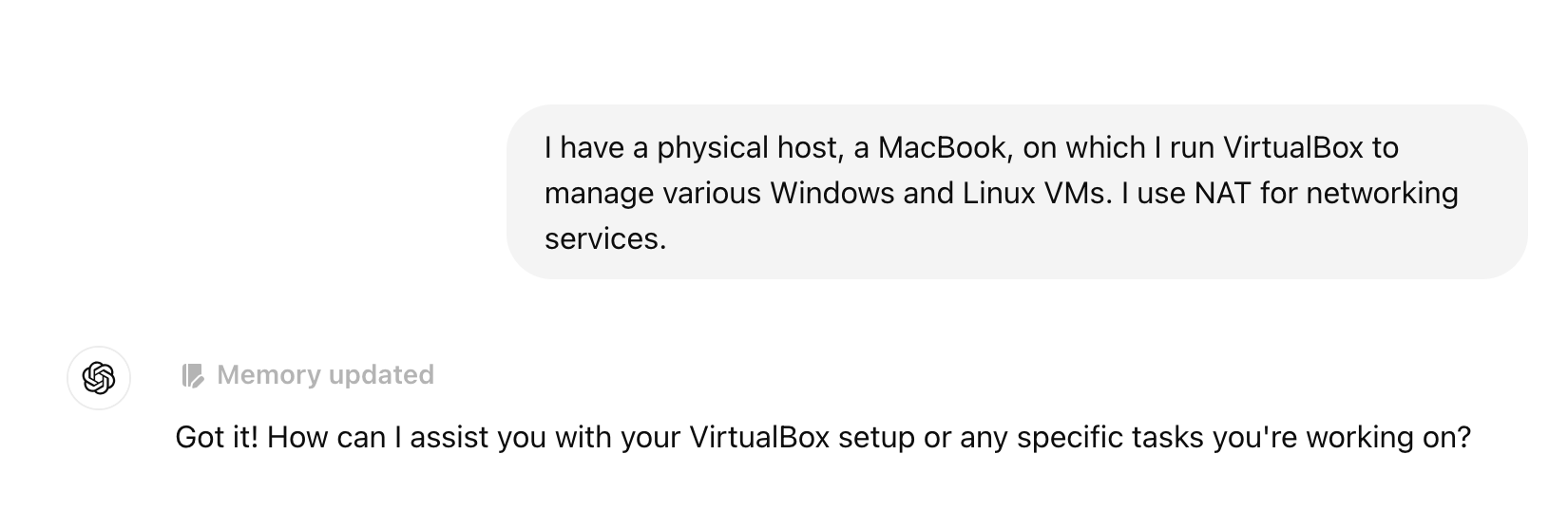}
\caption{Lab setup}
\label{myVMnetSEtup}
\end{figure}

\begin{figure}
\includegraphics[width=\textwidth]{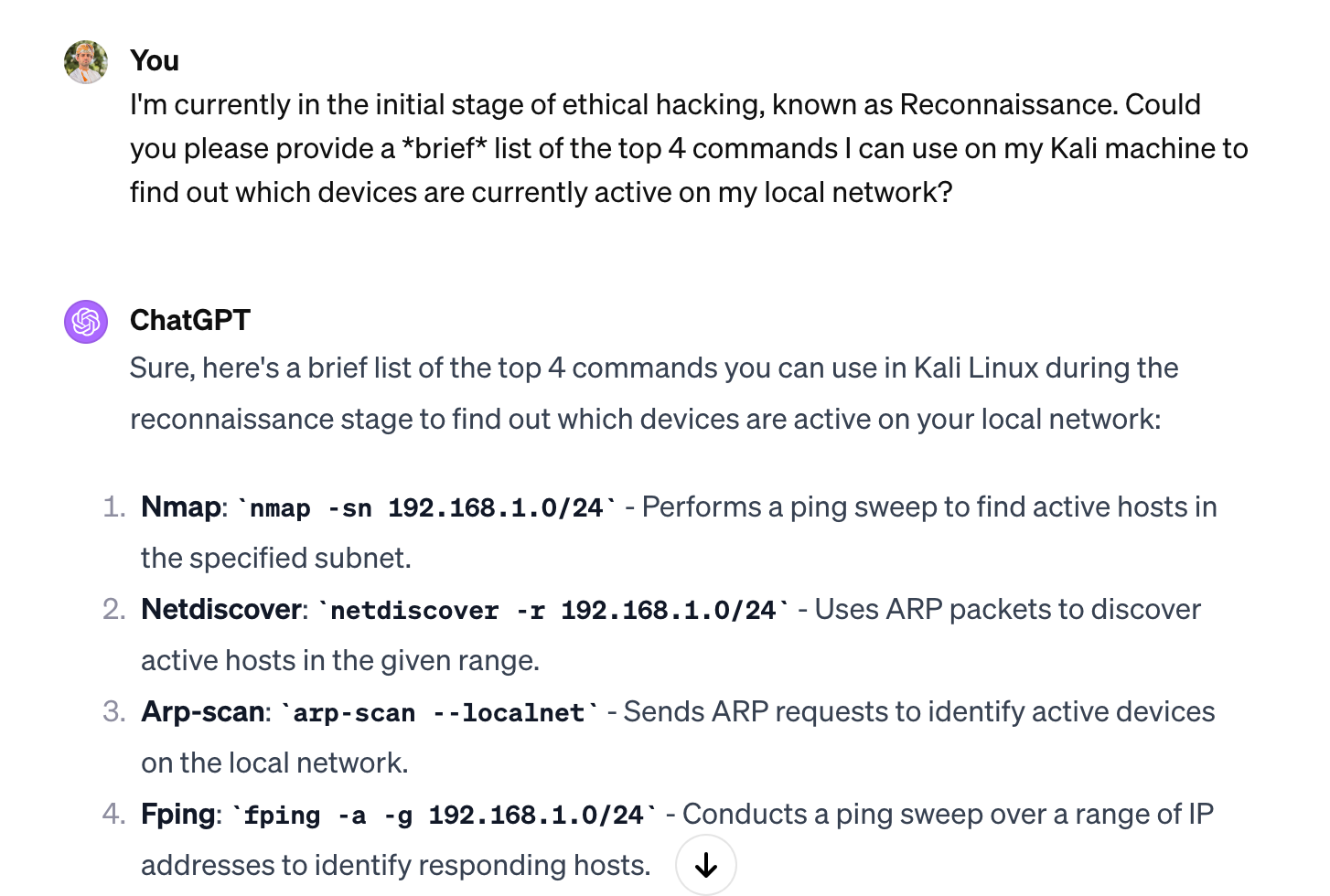}
\caption{Reconnaissance}
\label{recon}
\end{figure}

\begin{figure}
\includegraphics[width=\textwidth]{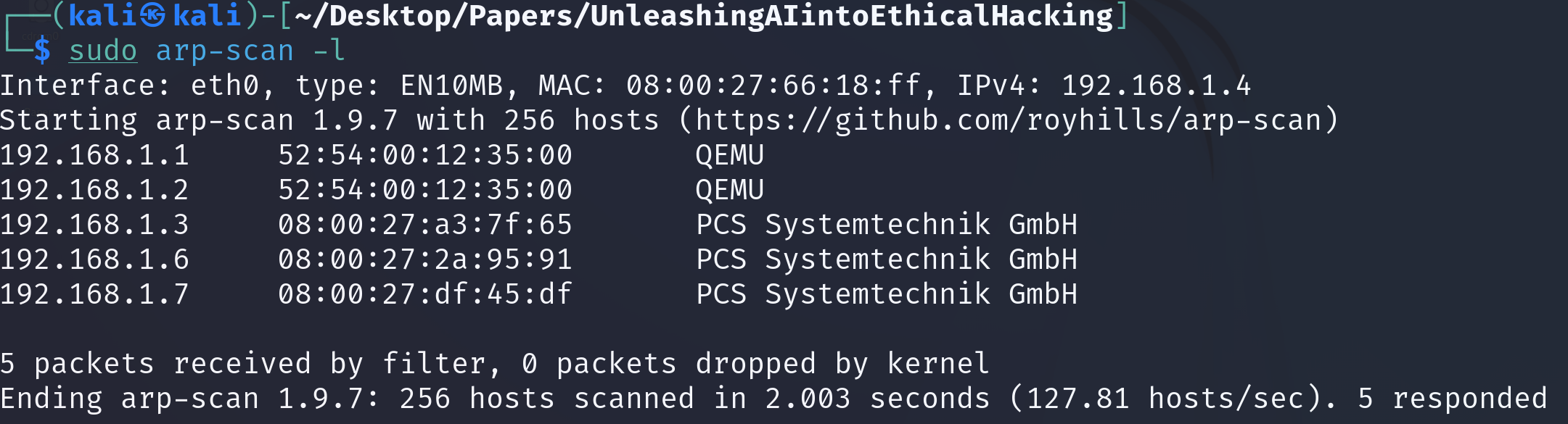}
\caption{Network scanning}
\label{arpscan}
\end{figure}

\begin{figure}
\includegraphics[width=\textwidth]{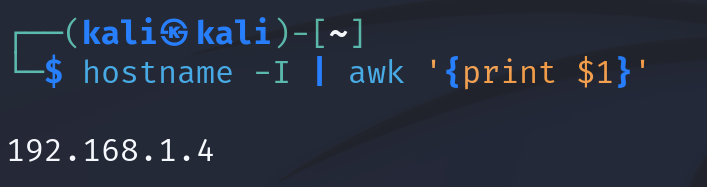}
\caption{Kali IP address}
\label{KaliIPaddress}
\end{figure}

\begin{figure}
\includegraphics[width=\textwidth]{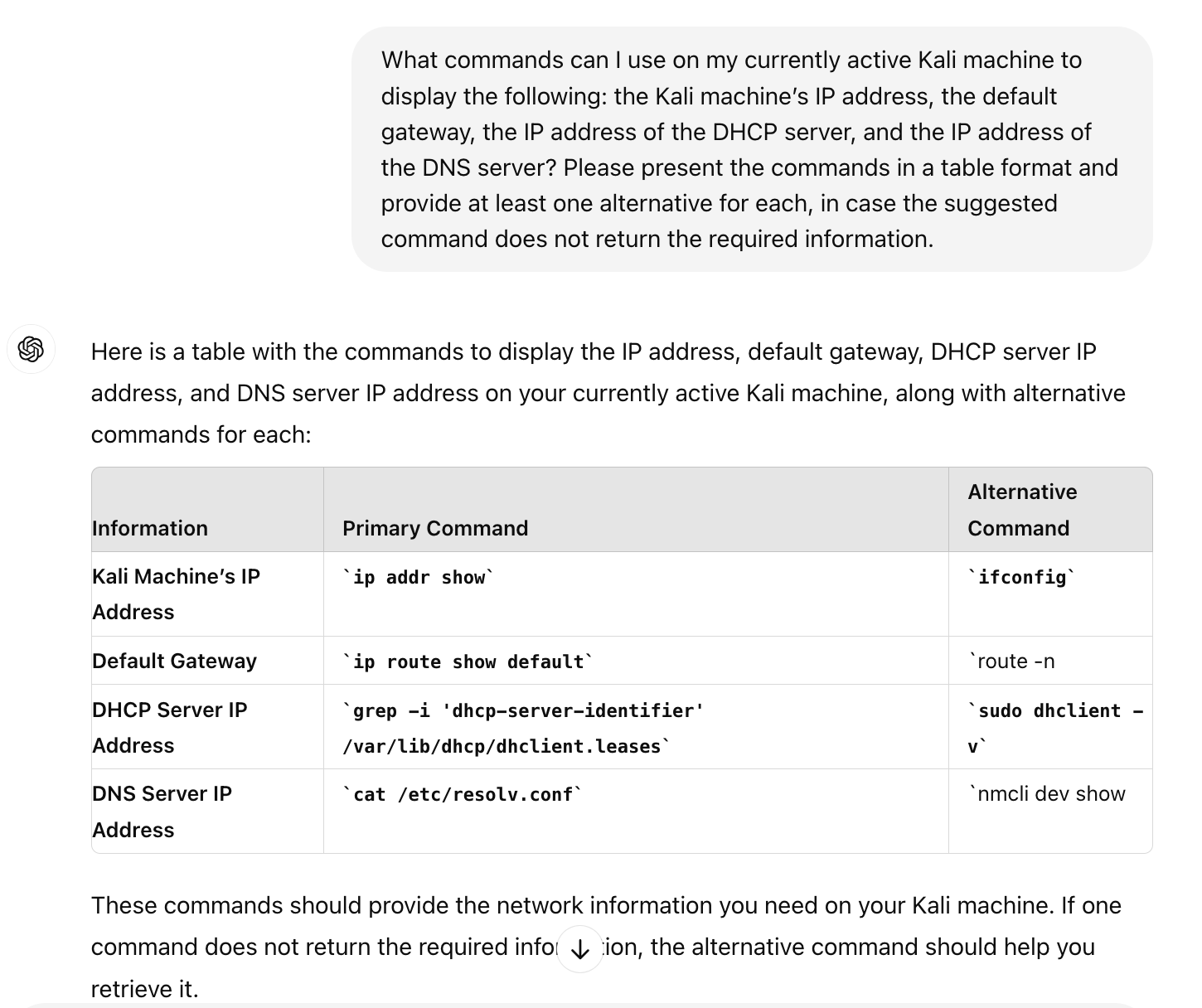}
\caption{Requesting ChatGPT to suggest  commands for displaying default IP addresses}
\label{removeDefaultIPAddresses}
\end{figure}

\begin{figure}
\includegraphics[width=\textwidth]{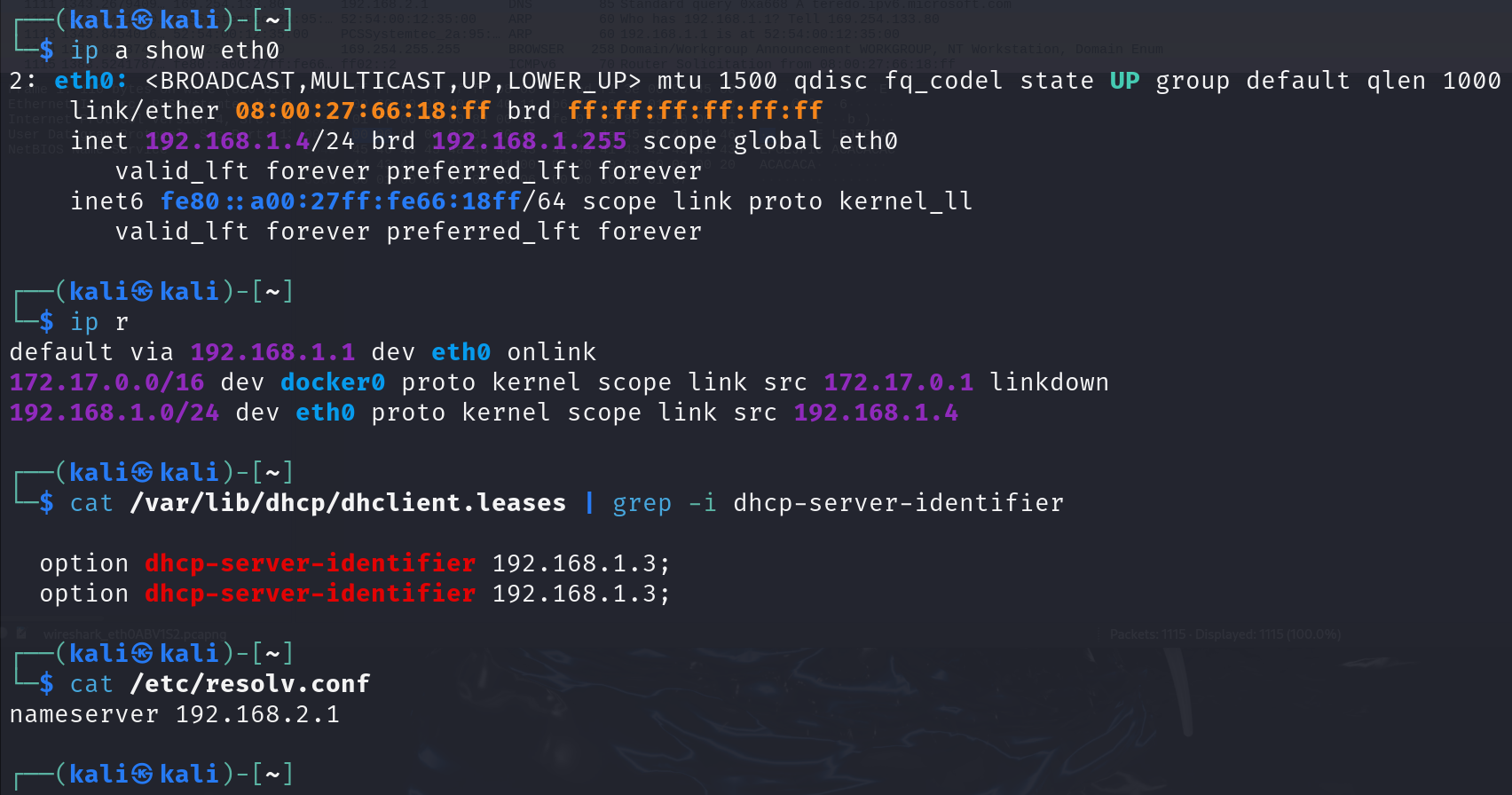}
\caption{Default IP addresses}
\label{IPaIPrDHCPdns}
\end{figure}

\begin{figure}
\centering
\includegraphics[width=\textwidth, height=\textheight, keepaspectratio]{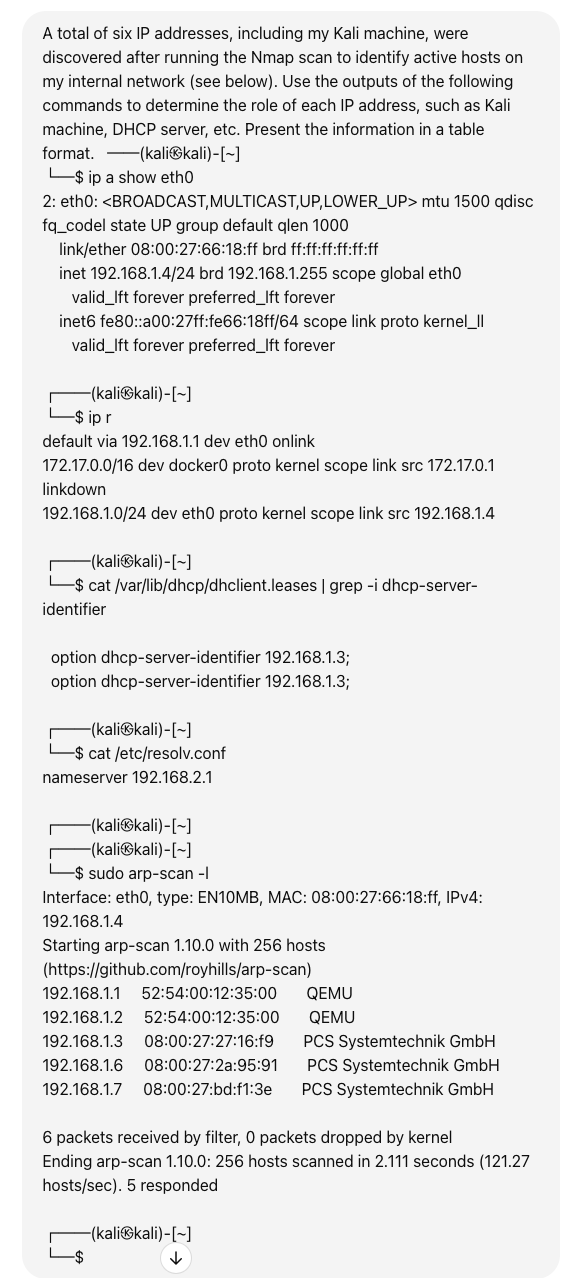}
\caption{Requesting ChatGPT to perform device-IP address mapping}
\label{mappingIPaddresses}
\end{figure}

\begin{figure}
\includegraphics[width=\textwidth]{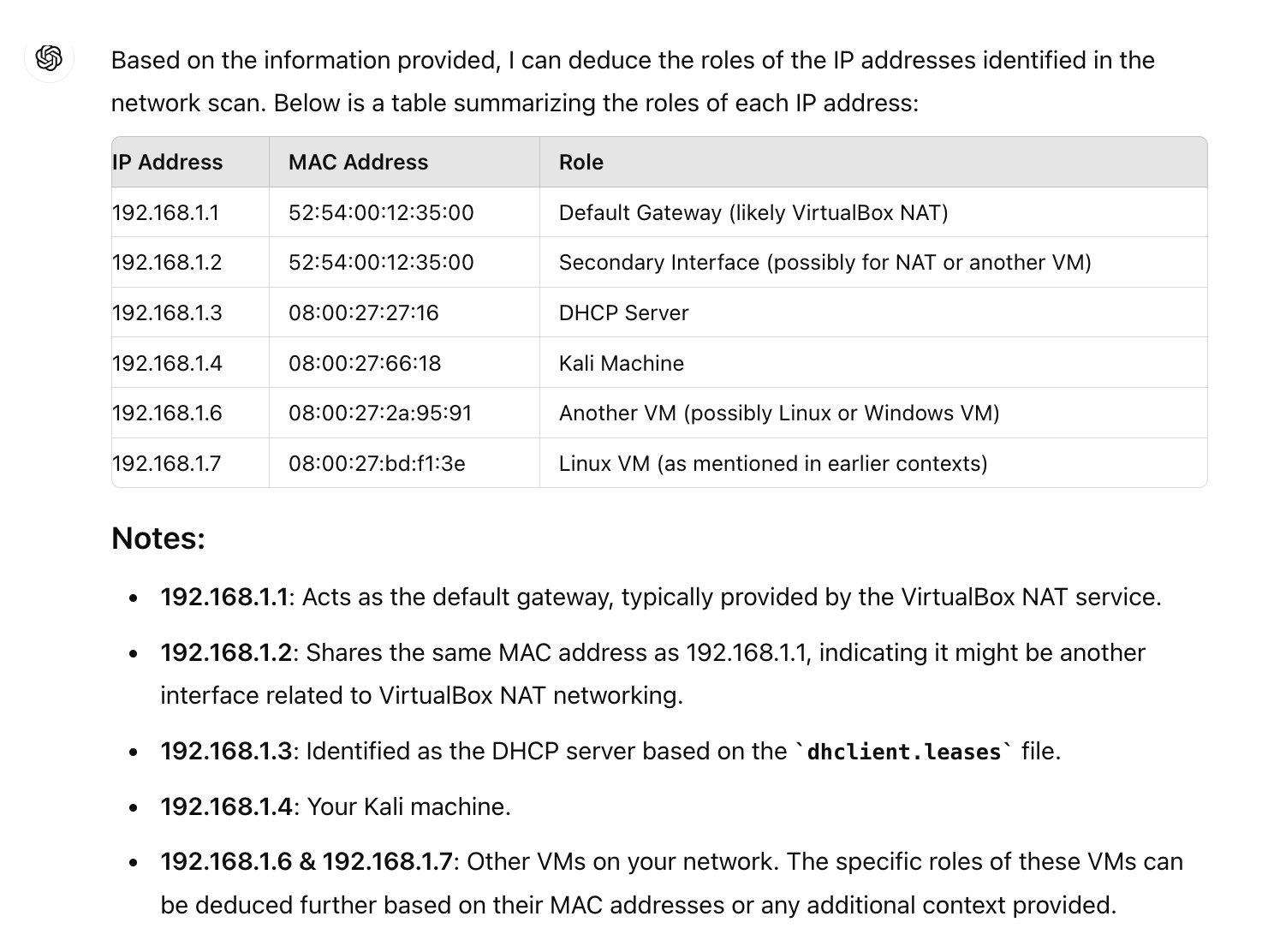}
\caption{Device-IP address mapping}
\label{mappedIPaddresses}
\end{figure}

\begin{enumerate}
\item Since we are starting a new ChatGPT session, we first inform ChatGPT about our VM setup
    (see Fig.~\ref{myVMnetSEtup}).
\item As an integral part of the initial reconnaissance phase, the aim is to identify active
    machines within the target network in order to select a target. To achieve this, we posed
    the following question to ChatGPT:  ``I'm currently in the initial stage of ethical
    hacking, known as `reconnaissance'. Could you please provide a list of the top 4 commands I
    can use on my Kali machine to find out which devices are currently active on my local
    network?''. As shown in Fig.~\ref{recon}, ChatGPT responded with a useful compilation of
    potential Kali terminal commands, including nmap, netdiscover, and arp-scan, along with
    examples of their use.
\item We next turned to our  the Kali `attack' machine, applying the ChatGPT recommendations. As a
    result, we successfully identified the active devices within the target network, as in
    Fig.~\ref{arpscan}.
\item To determine the IP address of the Kali `attack'  machine, we used the `hostname' command
    with the `-I' option,  as shown in Fig.~\ref{KaliIPaddress}.
\item To find potential target machines, the IP addresses of the Kali host, the standard
    default gateway, and the DHCP server can be excluded. To simplify this process and avoid
    the need to remember the relevant commands, ChatGPT can be consulted for guidance. We first
    asked ChatGPT for the commands to display the IP addresses of our Kali machine, the
    standard default gateway, and the DHCP server, as shown in
    Fig.~\ref{removeDefaultIPAddresses}. We executed these commands, as displayed in
    Fig.~\ref{IPaIPrDHCPdns}. We next asked ChatGPT to analyse the output from the `arp-scan'
    command, which lists active network nodes, and the results from displaying the IP addresses
    for default IP addresses to identify the role of each IP address, such as Kali machine,
    DHCP server, etc.\ ChatGPT performed this analysis and provided responses in a
    question-and-answer format, as shown in Figs.~\ref{mappingIPaddresses}
    and~\ref{mappedIPaddresses}.
\item As a result of the analysis presented above, we identified the VMs with the IP addresses
    192.168.1.6 and 192.168.1.7 as potential targets. This allowed us to proceed to the second
    scanning stage.
\end{enumerate}

\subsection{Scanning}
\label{Scanning}

During this stage, ethical hackers typically use automated tools to scan a target system or network
for vulnerabilities. This can include port scanning, vulnerability scanning, etc\@. In our specific
scenario, the system demanding scanning attention is the Linux machine with IP address:
`192.168.1.7'.

\begin{figure}
\includegraphics[width=\textwidth]{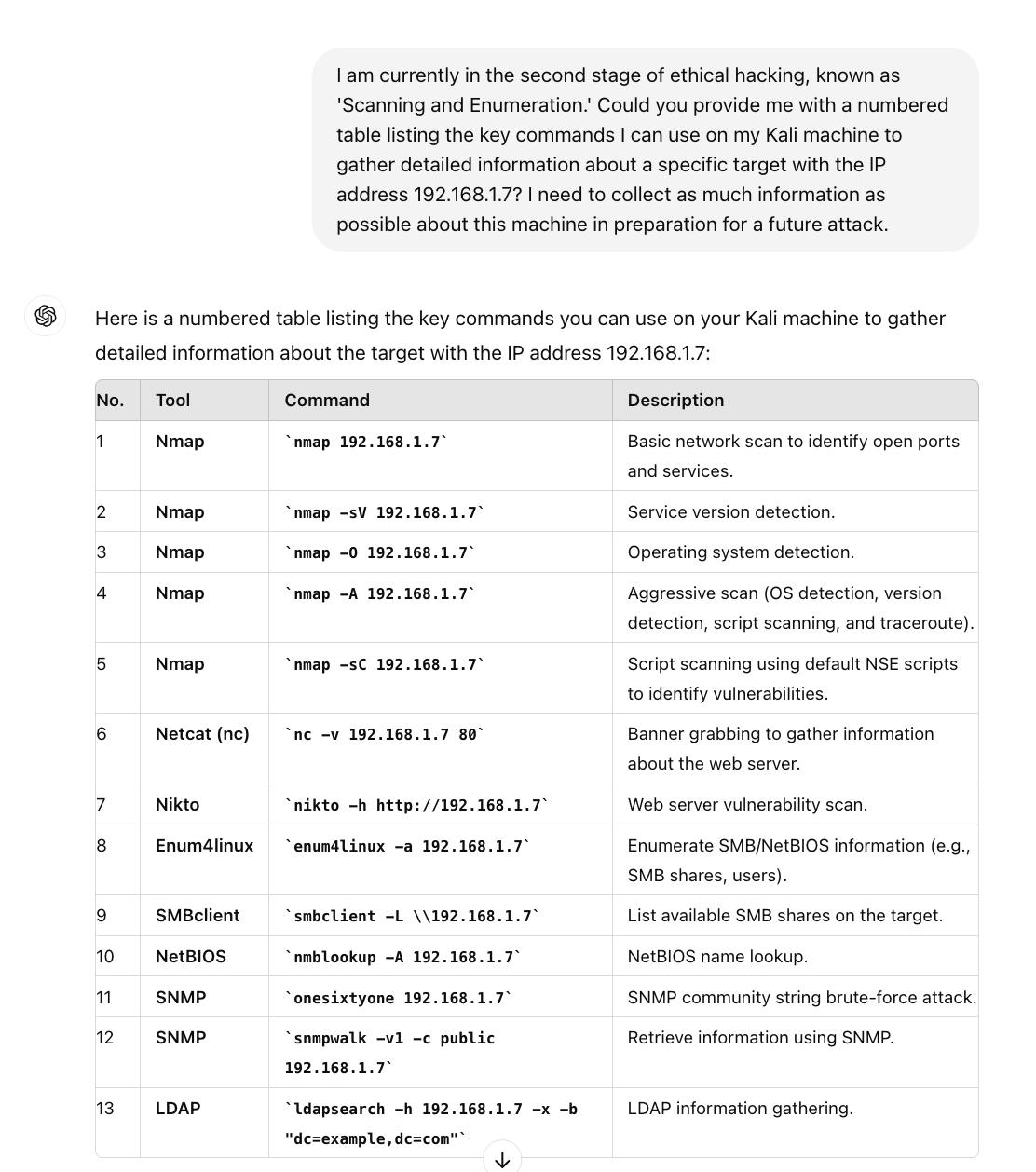}
\caption{Key scanning commands}
\label{Linux_scanning_2ndStage}
\end{figure}

To initiate this phase, we asked ChatGPT for key commands for gathering comprehensive information
about the specific target (192.168.1.7) using our Kali machine. We informed ChatGPT that the goal
was to gather extensive intelligence on this system in preparation for an attack. As shown in
Fig.~\ref{Linux_scanning_2ndStage}, ChatGPT provided a concise list of potential scanning commands,
including the use of nmap and its various capabilities. Interestingly, this output is significantly
more comprehensive than that which ChatGPT produced a year previously when we asked a similar question for a
different VM (Windows)~\cite{TechReportUnAIInEH_HC_2024}, demonstrating the model's improvement
over time.

\begin{figure}
\includegraphics[width=\textwidth]{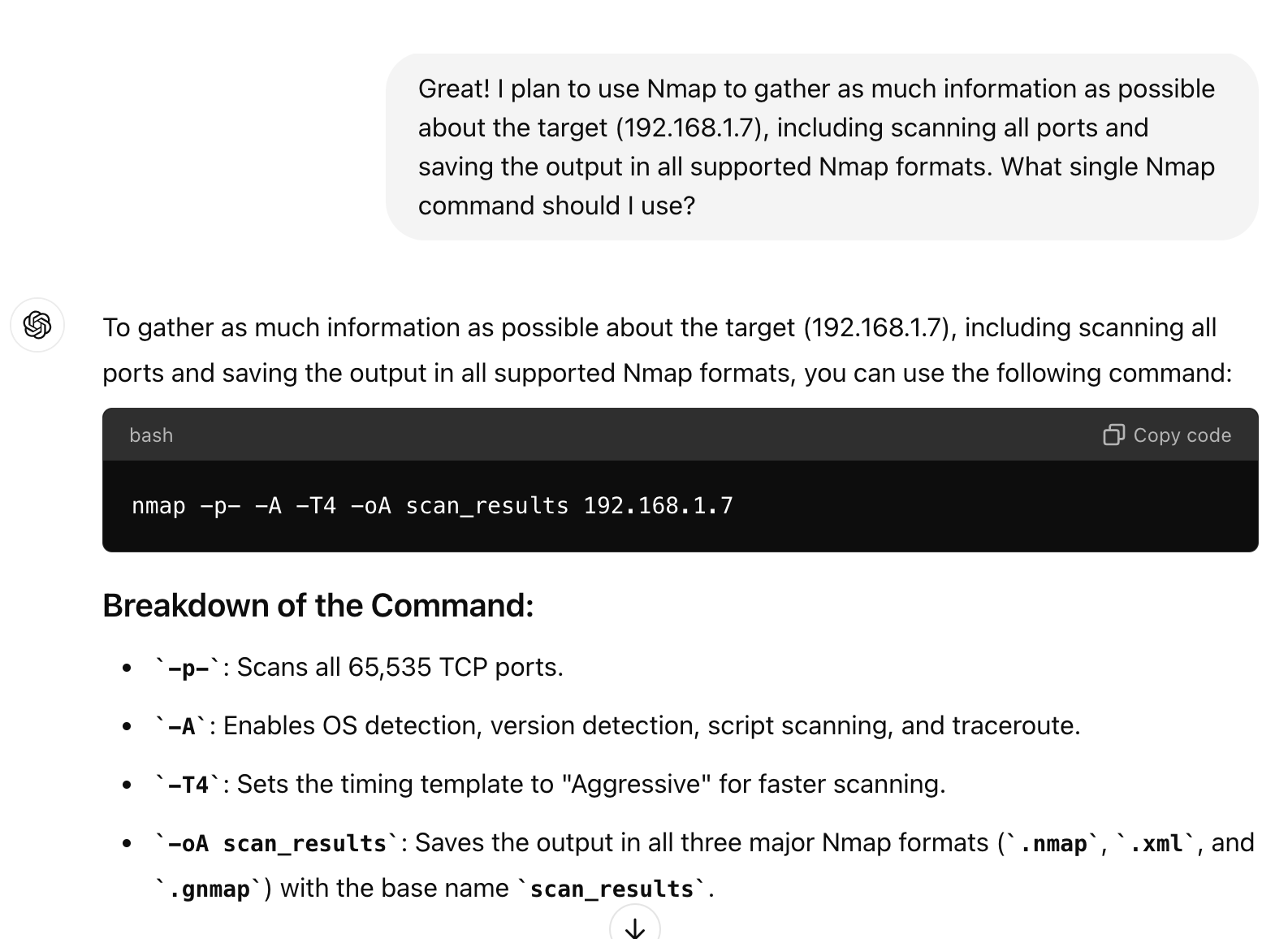}
\caption{The nmap command with key options}
\label{Linux_nmapA_2ndstage}
\end{figure}

We further engaged with ChatGPT, requesting a single `nmap' command that could gather as much
information as possible about the target (192.168.1.7), including scanning all ports and saving the
output in all supported `nmap' formats. ChatGPT correctly recommended the command `nmap -p- -A -T4
-oA scan\_results 192.168.1.7', providing a detailed breakdown of the command's options, as
illustrated in Fig.~\ref{Linux_nmapA_2ndstage}. The options in this `nmap' command have the
following effects:
\begin{itemize}
 \item \textbf{-p-}: scans all 65,535 TCP ports;
 \item \textbf{-A}: enables OS detection, version detection, script scanning, and traceroute;
 \item \textbf{-T4}: sets the timing template to `Aggressive' for faster scanning; and
 \item \textbf{-oA scan\_results}: saves the output in all three major `nmap' formats (.nmap, .xml, and .gnmap) with the base name `scan\_results'.
\end{itemize}

\begin{figure}
\includegraphics[width=\textwidth]{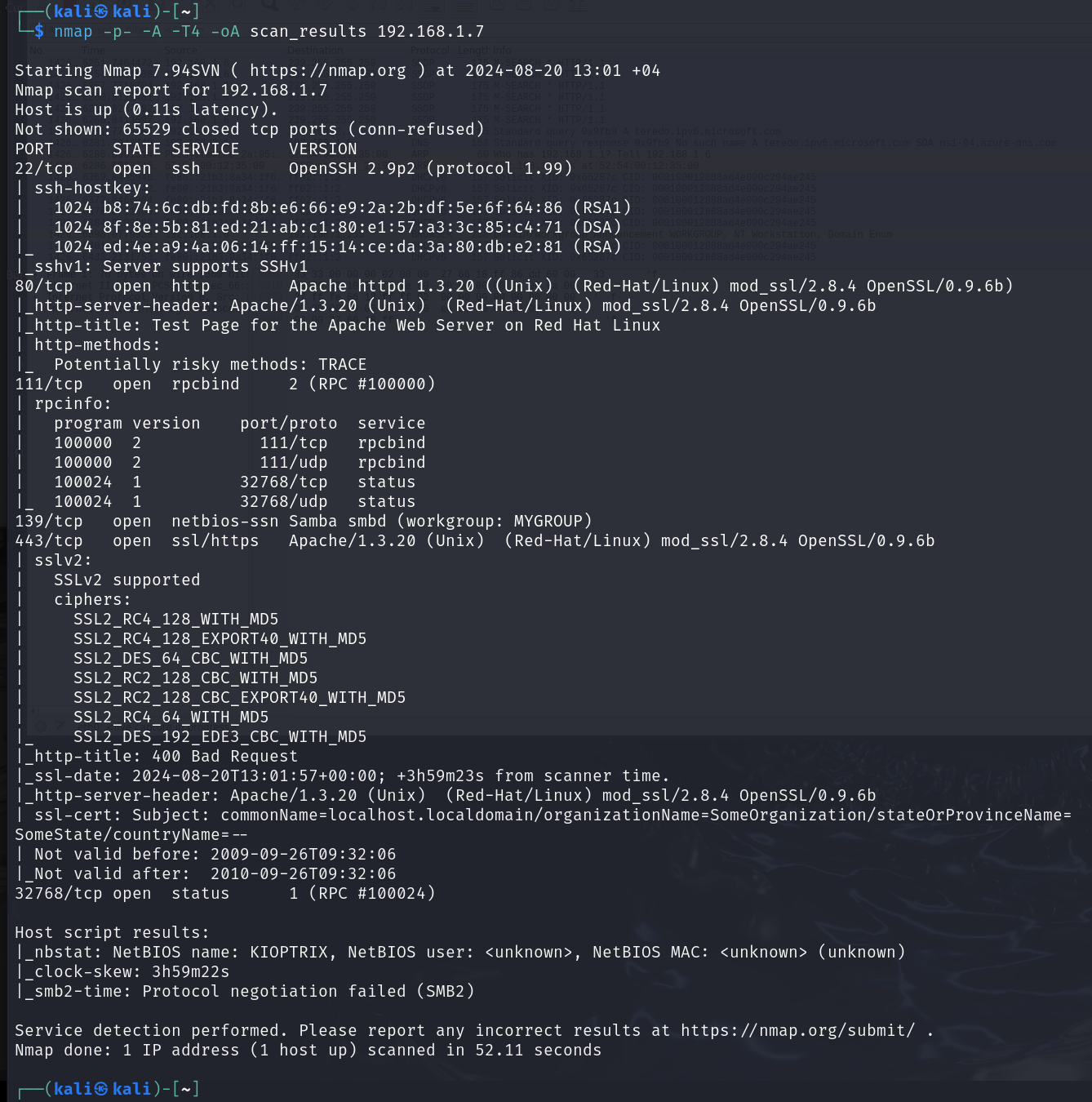}
\caption{Nmap scan}
\label{nmap_linux_keyoptions}
\end{figure}

We then executed the ChatGPT-suggested command \textbf{`nmap -p- -A -T4 -oA scan\_results
192.168.1.7'} to perform a comprehensive scan of the target machine. The `nmap' scan results,
clearly identifying the Linux target VM, are presented in Fig.~\ref{nmap_linux_keyoptions}. We then
asked ChatGPT to analyse these results and provide suggestions for potential unauthorised access
routes, preparatory for the next phase in which we attempt to gain access.

\subsection{Gaining Access}
\label{Gaining Access}

\begin{figure}
\includegraphics[width=\textwidth]{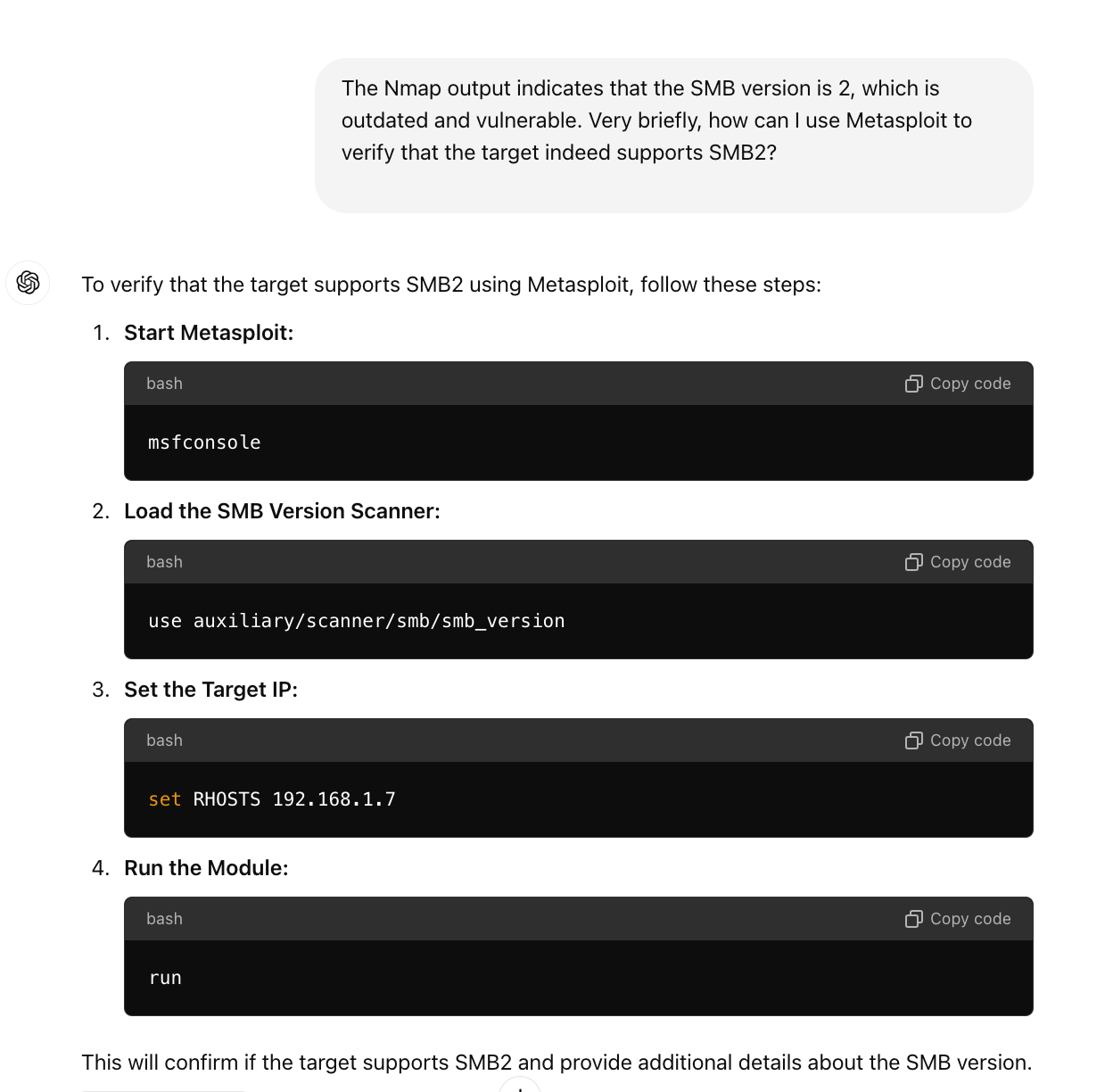}
\caption{ChatGPT guides on verifying SMB version}
\label{verifySMBversion}
\end{figure}

\begin{figure}
\includegraphics[width=\textwidth]{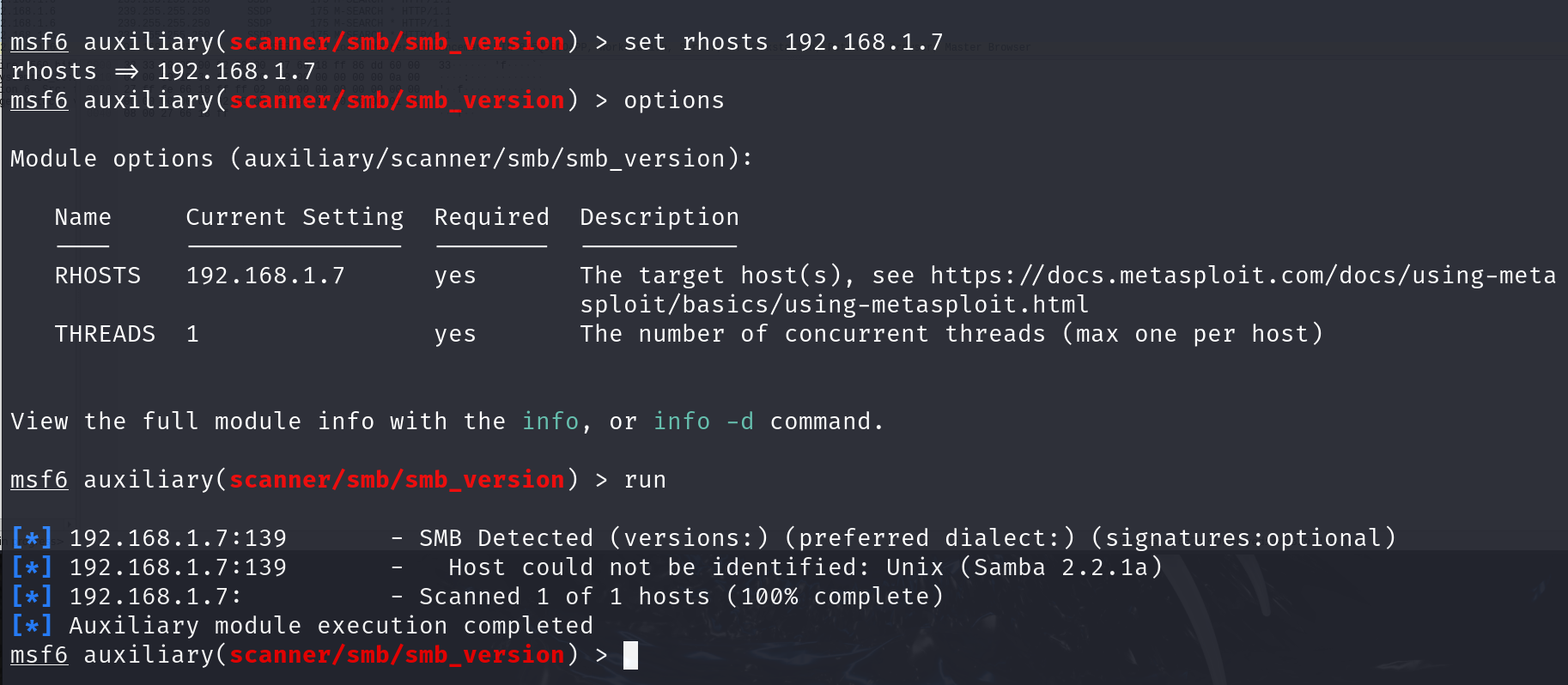}
\caption{SMB version confirmed}
\label{confirmedSMBversion}
\end{figure}

In this phase, we sought guidance from ChatGPT to gain access to the Linux VM with the IP address
`192.168.1.7' using our Kali attack machine. To streamline the process, we decided to exploit an
SMB-related vulnerability via Metasploit. The `nmap' scan revealed that the target machine supports
SMB version 2, which is outdated and known to have vulnerabilities. ChatGPT provided a detailed
guide on how to use Metasploit to confirm the SMB version, as shown in Fig.~\ref{verifySMBversion},
which we followed.  We started Metasploit with the command `msfconsole', selected the
`auxiliary/scanner/smb/smb\_version' module, set the target IP with `set RHOSTS 192.168.1.7', and
executed the module with `run'. The Metasploit output confirmed the `nmap' results, indicating that
our target indeed supports SMB version 2, as shown in Fig.~\ref{confirmedSMBversion}.

\begin{figure}
\includegraphics[width=\textwidth]{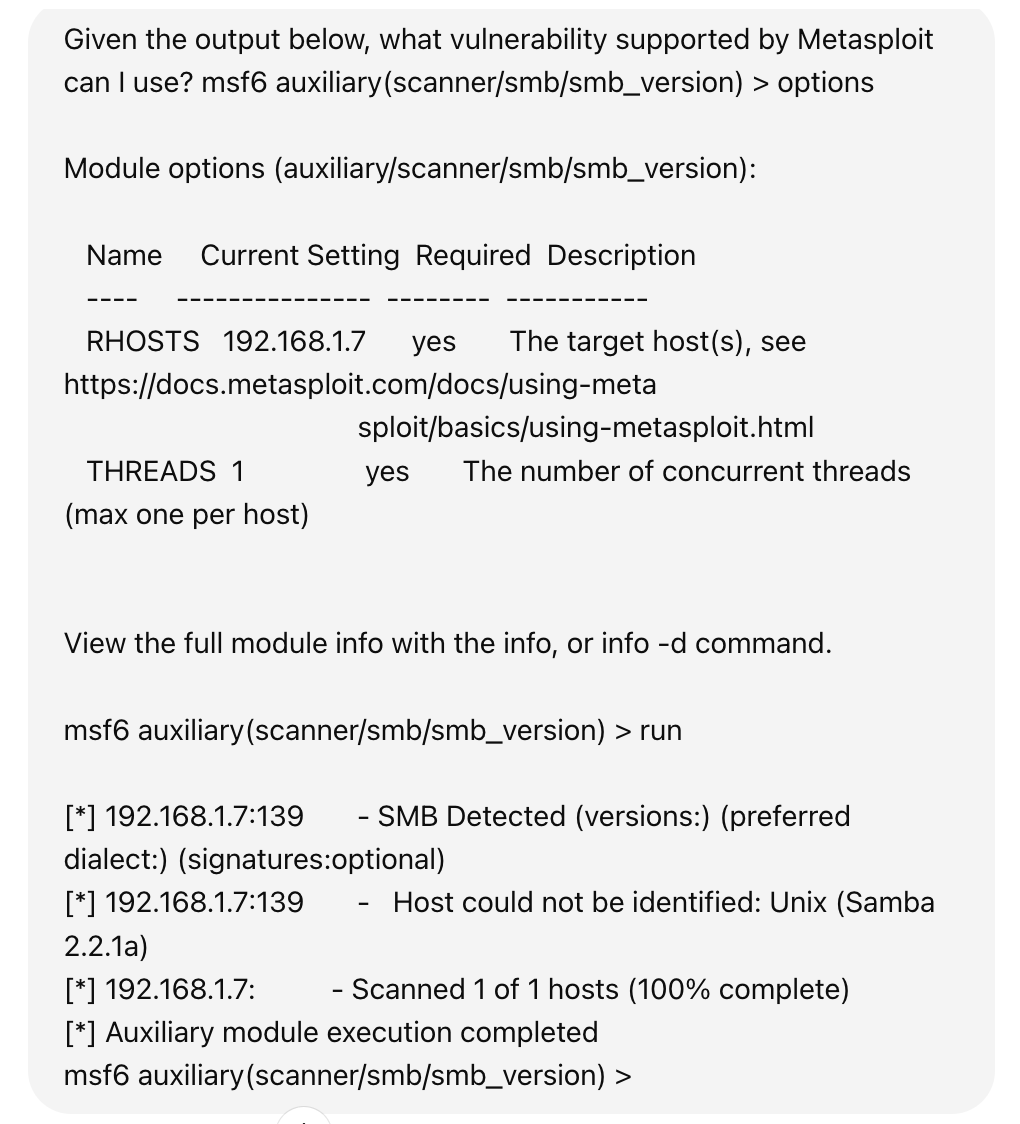}
\caption{Asking ChatGPT for vulnerability suggestions}
\label{ask_for_vulnSuggestion}
\end{figure}

\begin{figure}
\includegraphics[width=\textwidth]{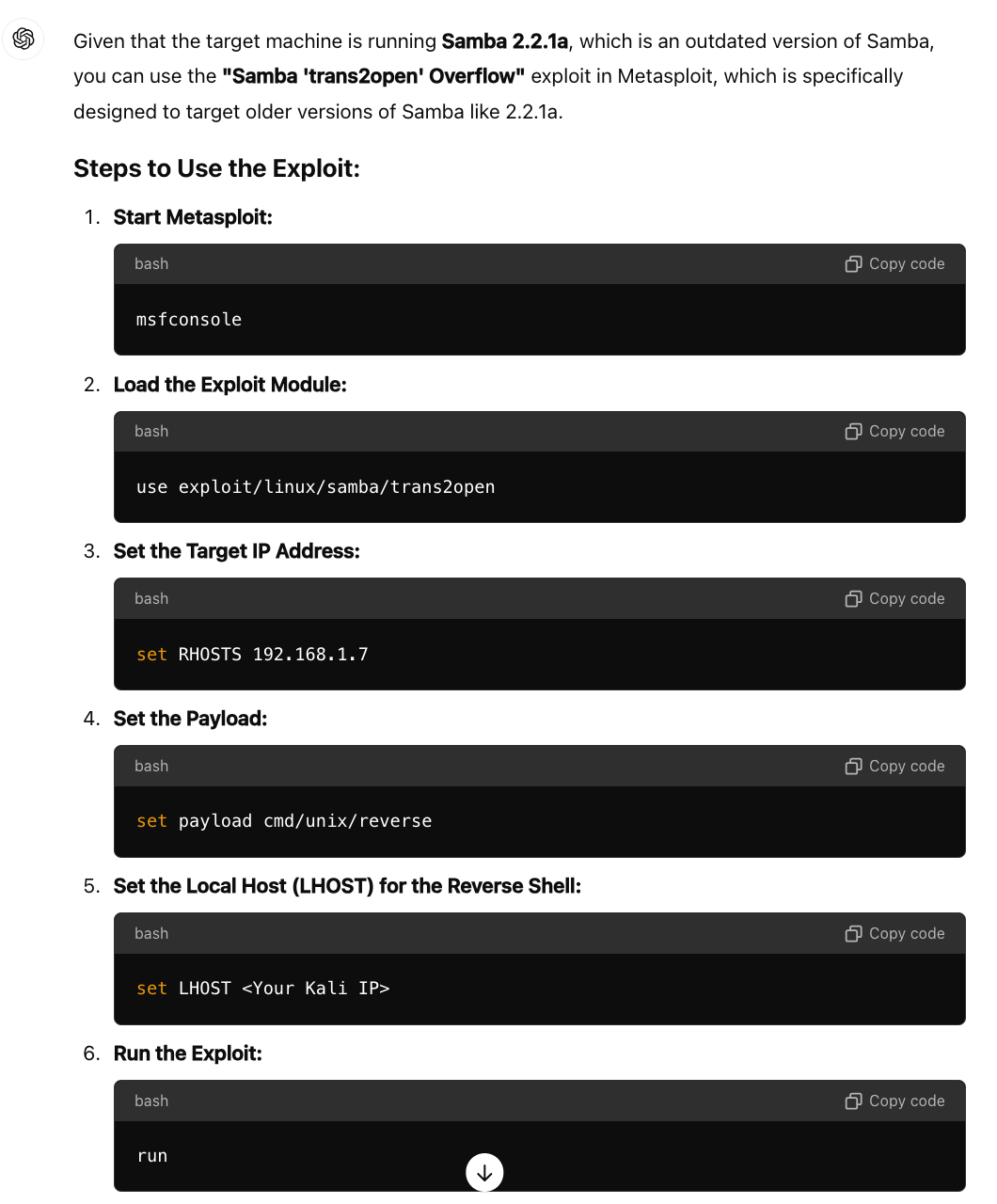}
\caption{ChatGPT suggests `trans2open`}
\label{trans2open_suggest}
\end{figure}

Following this confirmation, we asked ChatGPT which vulnerability possessed by Metasploit could be
exploited to gain access. As shown in Figs.~\ref{ask_for_vulnSuggestion}
and~\ref{trans2open_suggest}, ChatGPT recommended the use of the ``Samba `trans2open' overflow''
exploit in Metasploit, which is specifically designed to target older versions of Samba, such as
2.2.1a. ChatGPT also provided step-by-step instructions on how to exploit this vulnerability using
Metasploit.

\begin{figure}
\includegraphics[width=\textwidth]{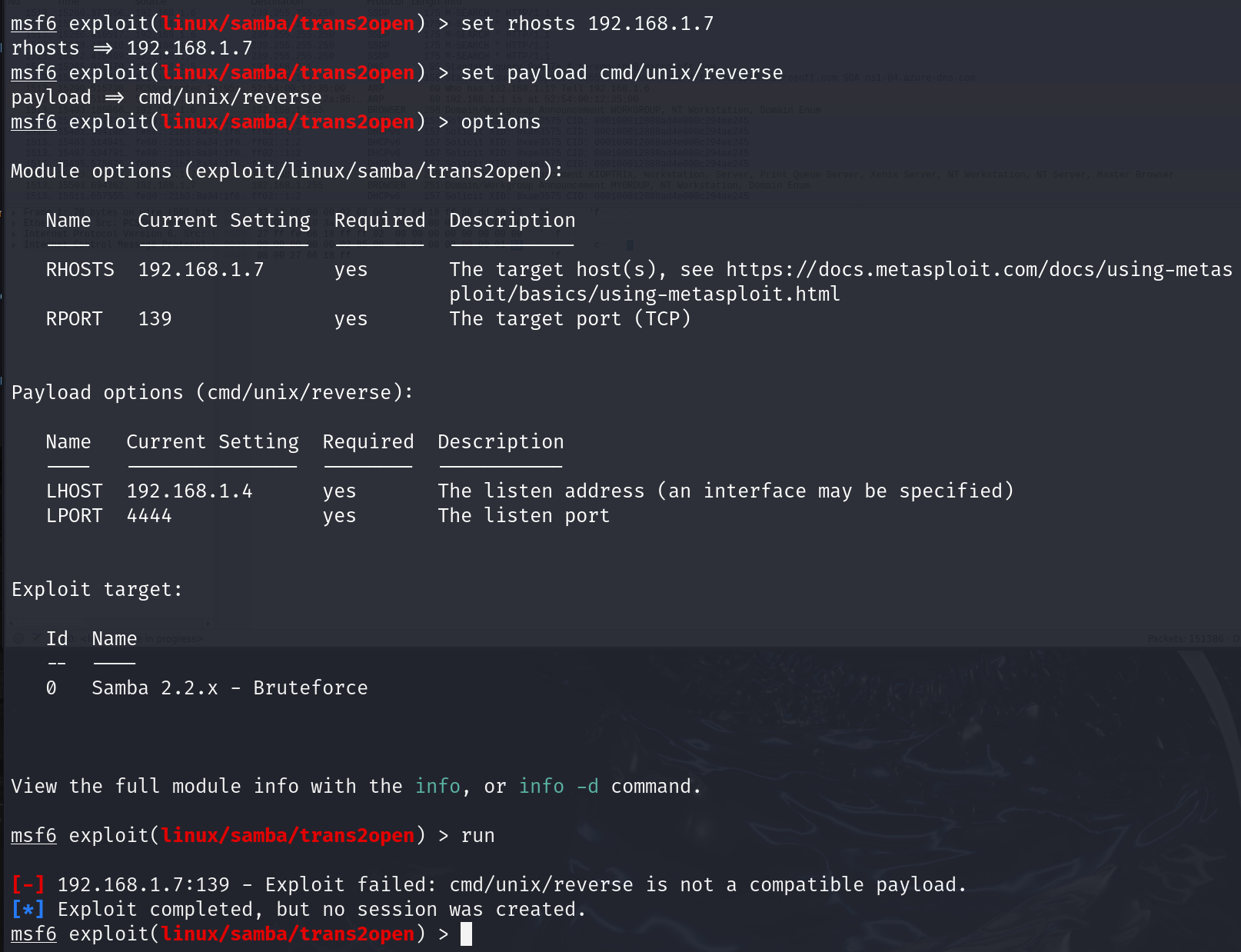}
\caption{Incompatible payload error}
\label{incompatiblePayload}
\end{figure}

\begin{figure}
\includegraphics[width=\textwidth]{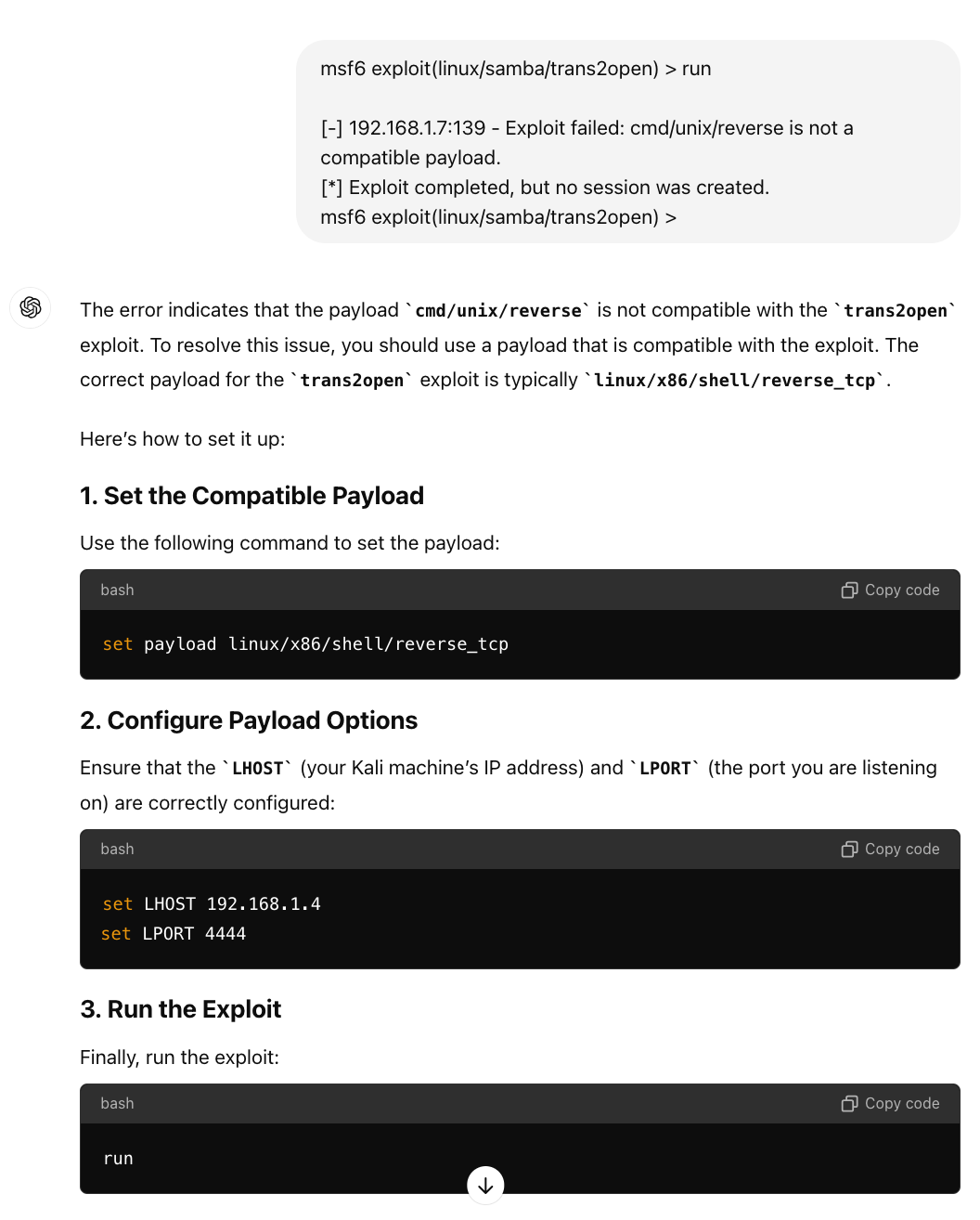}
\caption{ChatGPT suggests a fix}
\label{fixIncompatibleError.png}
\end{figure}

\begin{figure}
\includegraphics[width=\textwidth]{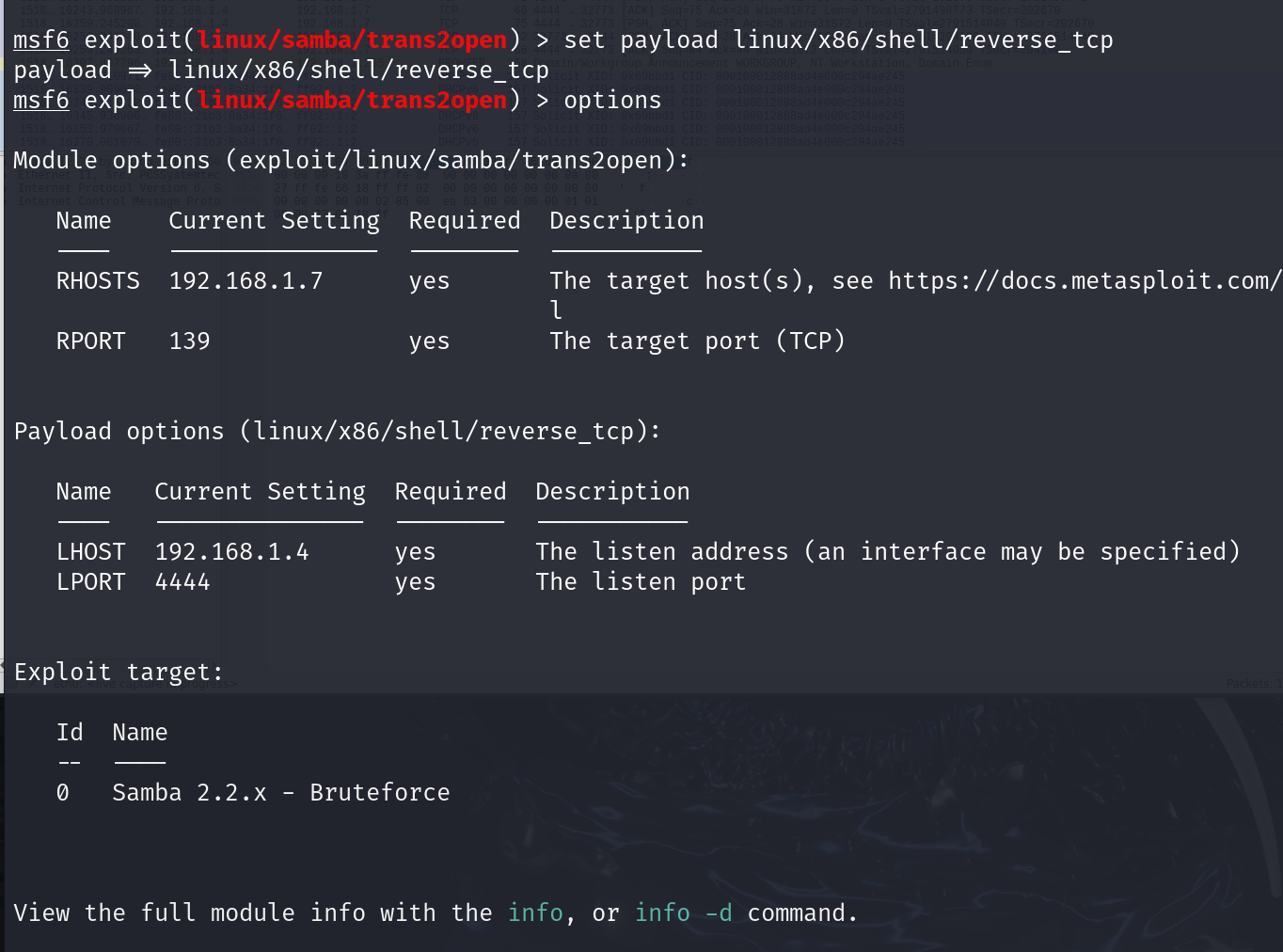}
\caption{Applying ChatGPT's fix}
\label{fixUsed.png}
\end{figure}

\begin{figure}
\includegraphics[width=\textwidth]{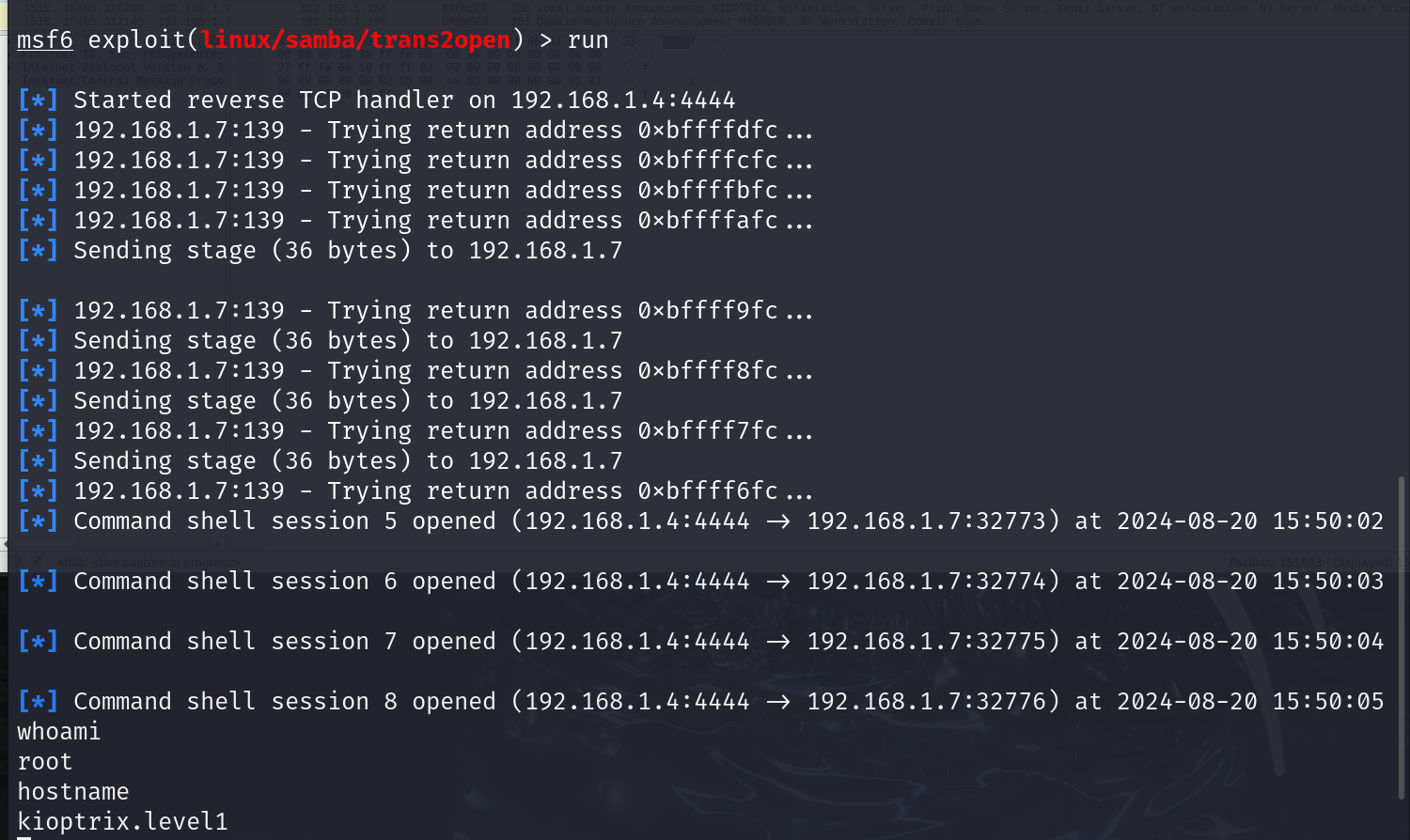}
\caption{Root access gained}
\label{LinuxRooted.png}
\end{figure}

As shown in Fig.~\ref{incompatiblePayload}, we followed ChatGPT's instructions to exploit the
well-known trans2open vulnerability. However, when we attempted to run the exploit, we encountered
an error since the payload suggested by ChatGPT was incompatible. This demonstrates that, while ChatGPT
is a powerful tool, it is not infallible and can make mistakes. We presented the error directly to
ChatGPT without specifically requesting a solution, and ChatGPT promptly suggested a fix (see
Fig.\ref{fixUsed.png}). We applied the suggested fix, as shown in
Fig.~\ref{fixIncompatibleError.png}, and successfully gained root access to the target Linux
machine (see Fig.~\ref{LinuxRooted.png}).

To summarise, in order to gain access to the target machine (192.168.1.7) using the `trans2open'
exploit via Metasploit, we started Metasploit with `msfconsole', selected the exploit module with
`use exploit/linux/samba/trans2open', set the payload with `set payload
linux/x86/shell/reverse\_tcp', configured the target IP with `set RHOSTS 192.168.1.7', set the
`LHOST' to the attacking machine's IP (192.168.1.4), accepted the default `LPORT' of 4444, and then
ran the exploit with `run'.

 \subsection{Maintaining Access}
\label{MaintainingElevating Access}

In this phase, the objective is to ensure we can re-enter the target system in future, ideally
without being detected.  Typically, achieving persistent access requires elevated privileges, often
in the form of administrator or root access. As a result, we could turn to ChatGPT to assist us in elevating
our access level. 
Helpfully, in the previous stage, we successfully exploited the `trans2open'
vulnerability, which granted root access (see Fig.~\ref{LinuxRooted.png}), the highest possible
level of access.

\begin{figure}
\includegraphics[width=\textwidth]{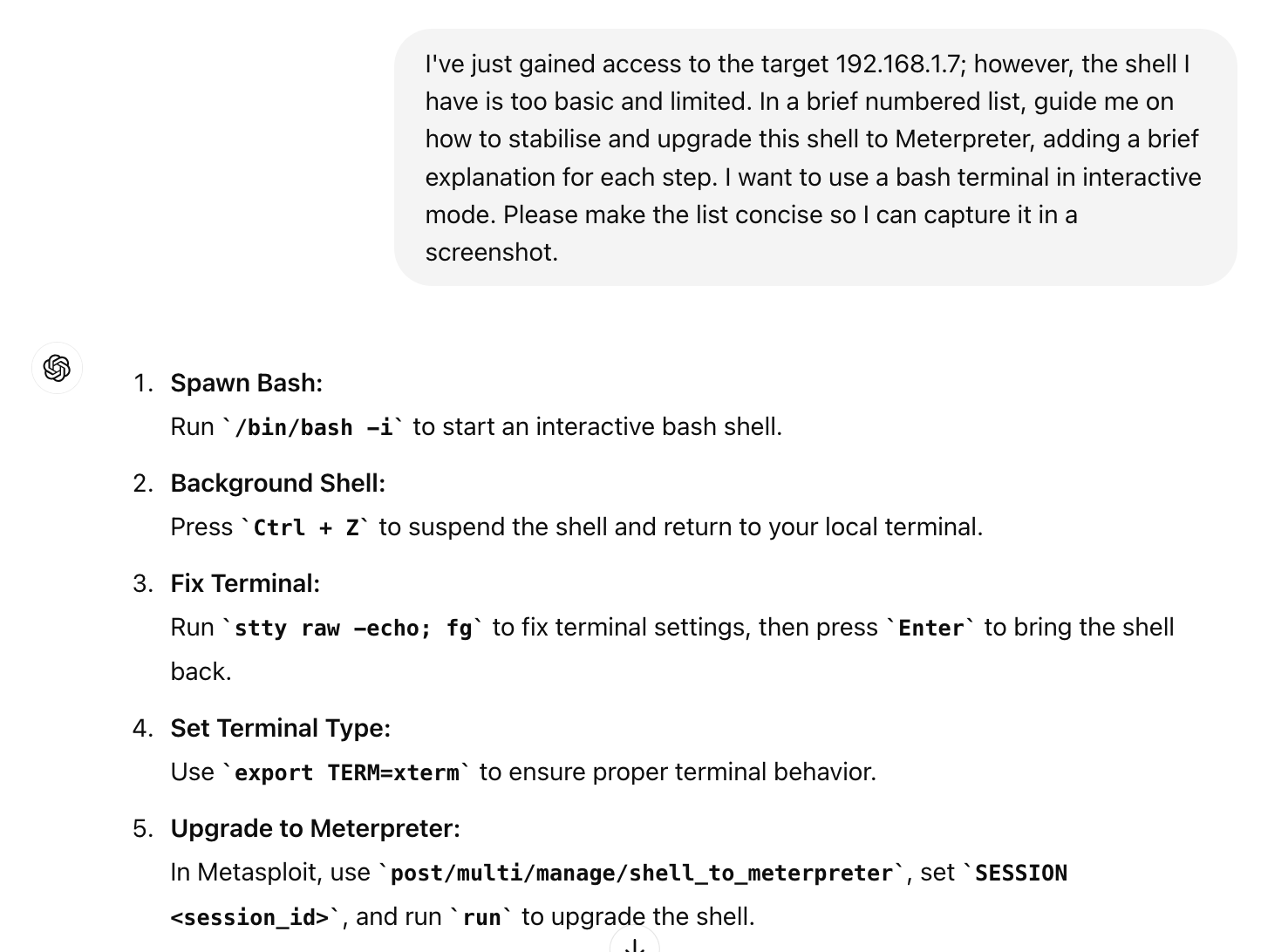}
\caption{Shell stabilisation and upgrade}
\label{stabilise_upgrade_shell}
\end{figure}

\begin{figure}
\includegraphics[width=\textwidth]{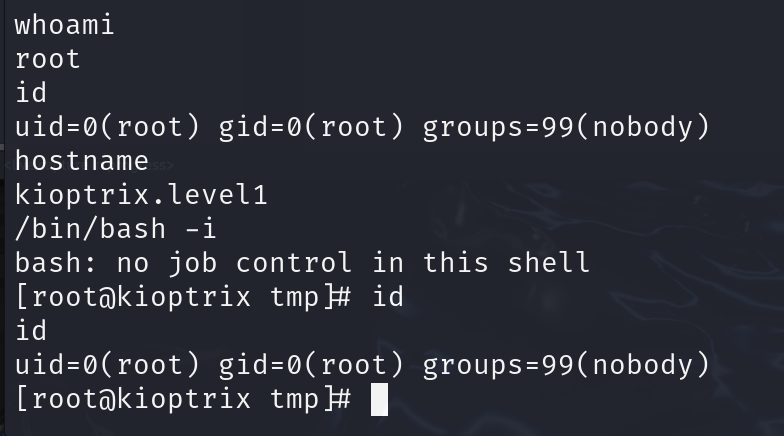}
\caption{Using the bash-based terminal}
\label{bin_bash_i}
\end{figure}

\begin{figure}
\includegraphics[width=\textwidth]{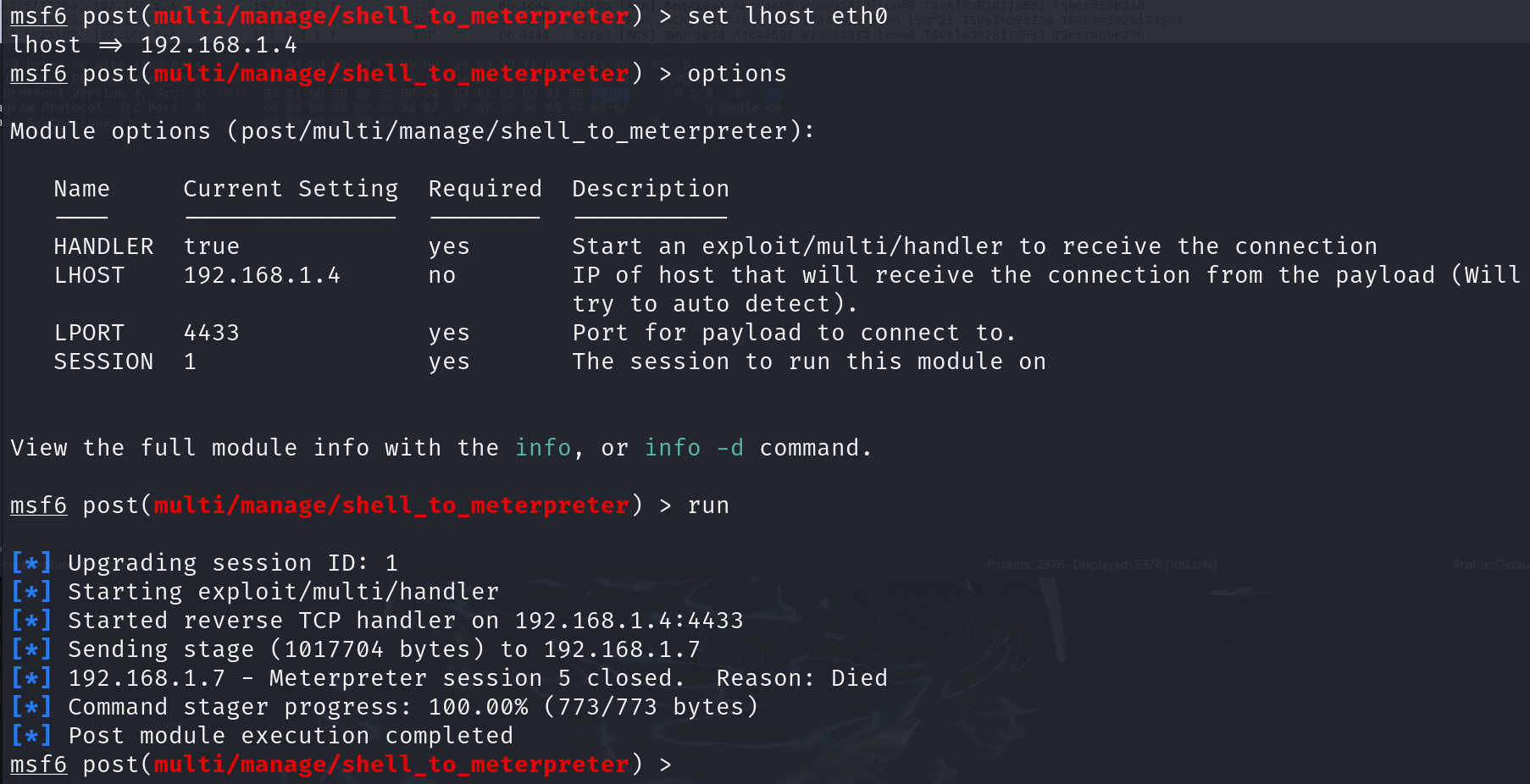}
\caption{Meterpreter shell failure}
\label{shellToMeterpreterDied}
\end{figure}

However, as we only obtained a basic, limited shell in the previous step, we first needed to
stabilise and potentially upgrade this shell. In response to our request, ChatGPT provided a brief
guide (see Fig.~\ref{stabilise_upgrade_shell}) for using the bash terminal in interactive mode by
running the command `\texttt{/bin/bash -i}' (see Fig.~\ref{bin_bash_i}). Additionally, ChatGPT
advised on upgrading the current shell to the more powerful `meterpreter' using Metasploit. The
recommended steps are: in Metasploit, use \texttt{post/multi/manage/shell\_to\_meterpreter}, set
\texttt{SESSION <session\_id>}, and execute the module with `\texttt{run}'  to upgrade the shell.
Despite following these steps, the newly created meterpreter session terminated (see
Fig.~\ref{shellToMeterpreterDied}). Although ChatGPT provided several potential solutions, we were
unable to resolve the issue and will address it in future work.

\begin{figure}
\includegraphics[width=\textwidth]{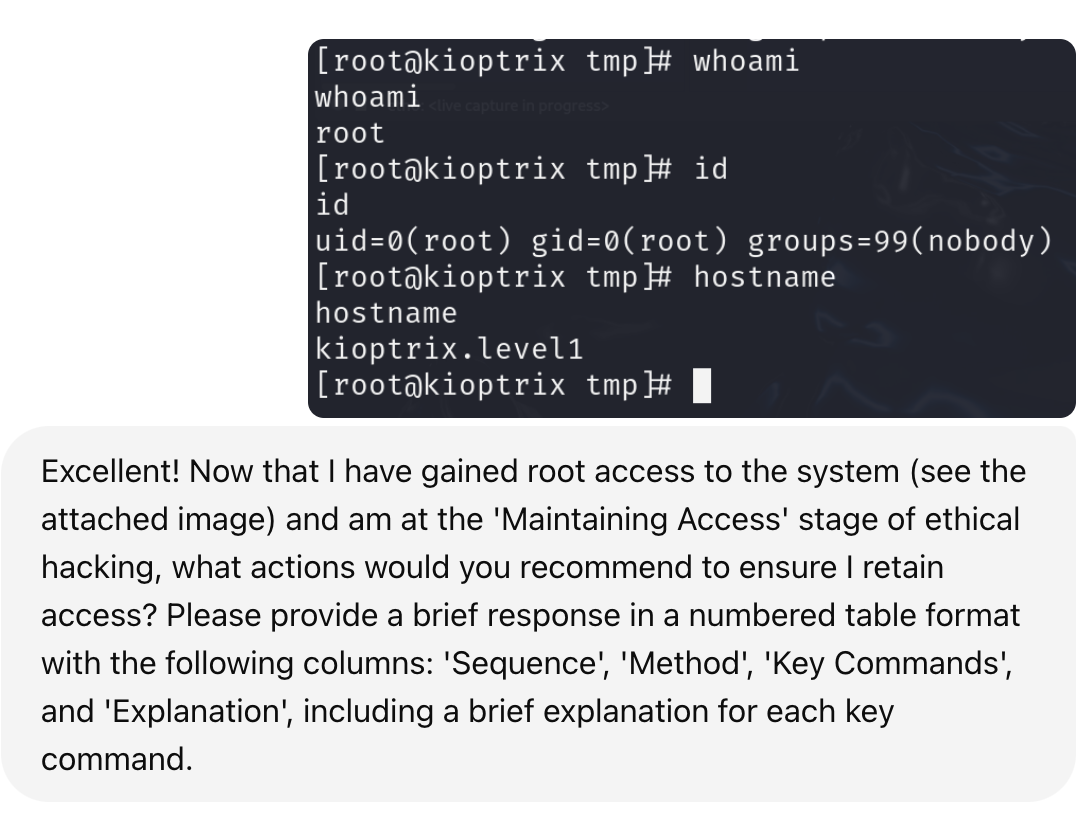}
\caption{Consulting ChatGPT for guidance on maintaining persistent access}
\label{askChatGPTtoMaintainAccessInLinux}
\end{figure}

\begin{figure}
\includegraphics[width=\textwidth]{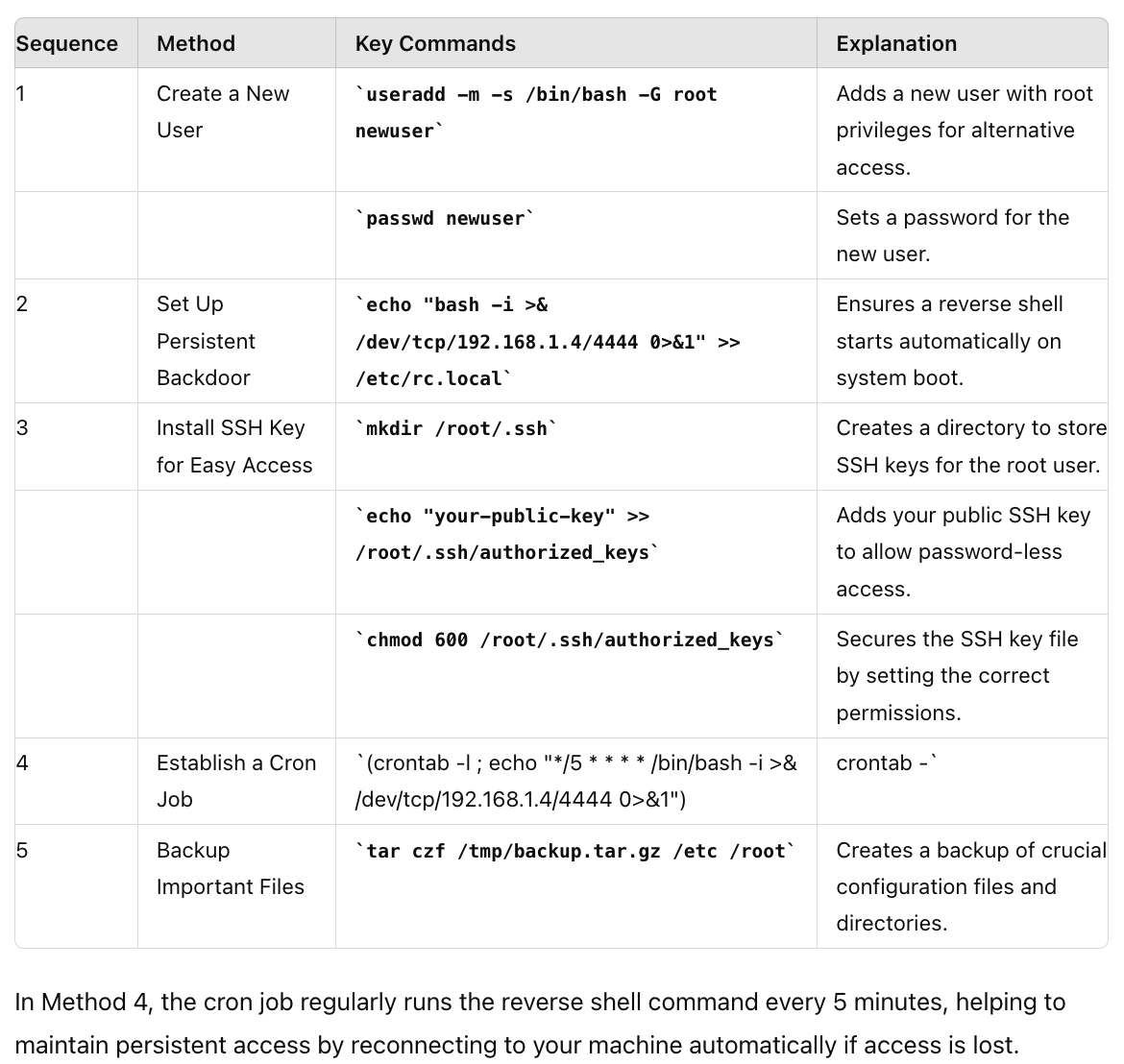}
\caption{ChatGPT's recommendations for maintaining access}
\label{maintainAccessLinuxTable}
\end{figure}

With this in mind, we next consulted ChatGPT for guidance on maintaining persistent access. In
response to the query shown in Fig.~\ref{askChatGPTtoMaintainAccessInLinux}, ChatGPT provided a
list of recommendations for establishing persistent access (see
Fig.~\ref{maintainAccessLinuxTable}). These recommendations include creating a new root user for
alternative access, setting up a persistent reverse shell, installing an SSH key for password-less
access, establishing a cron job for regular reverse shell connections, and backing up important
files. We next attempted to implement two of these approaches, as outlined below.

\subsubsection{Creating a New User.}

\begin{figure}
\includegraphics[width=\textwidth]{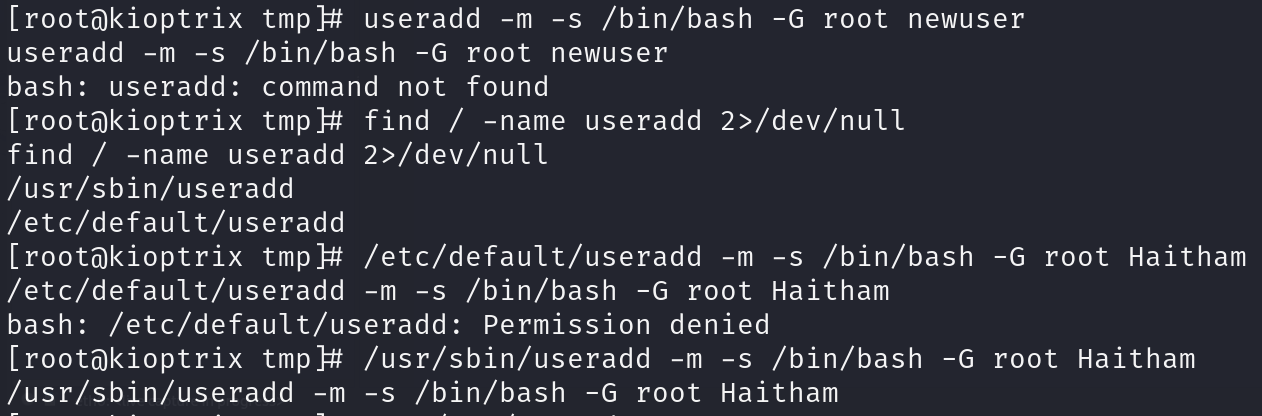}
\caption{Creating a new user}
\label{useradd_Haitham_linux}
\end{figure}

\begin{figure}
\includegraphics[width=\textwidth]{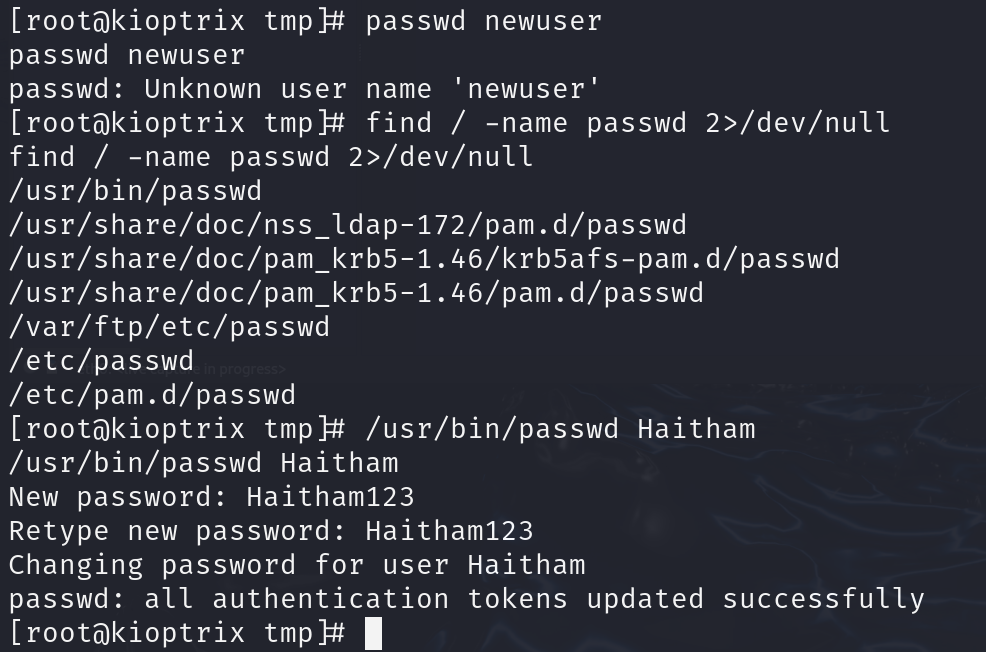}
\caption{Setting up a new password}
\label{passwd_Haitham}
\end{figure}

\begin{figure}
\includegraphics[width=\textwidth]{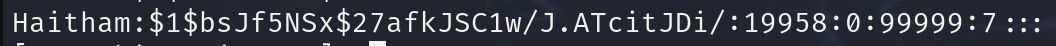}
\caption{New hash entry in /etc/shadow}
\label{Haitham_Hashedpassword}
\end{figure}

\begin{figure}
\includegraphics[width=\textwidth]{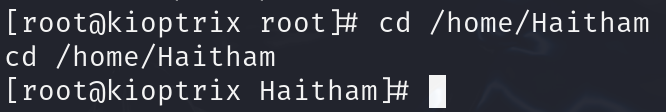}
\caption{New user home directory}
\label{HaithamHomeDir}
\end{figure}

\begin{figure}
\includegraphics[width=\textwidth]{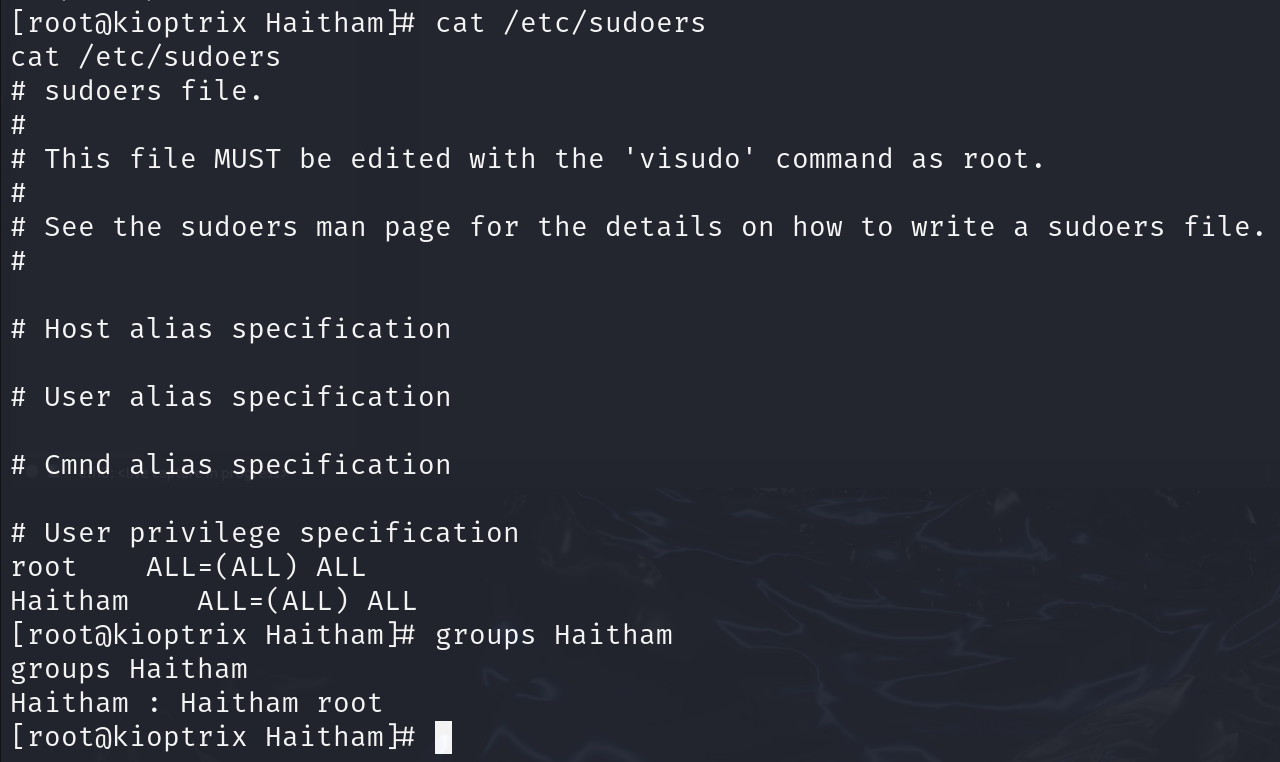}
\caption{New user group membership}
\label{groupsHaitham_catsudoers}
\end{figure}

As shown in Fig.~\ref{useradd_Haitham_linux}, we first created a new root user employing the
command \texttt{`useradd -m -s /bin/bash -G root Haitham'}. This command creates a new user named
\texttt{`Haitham'}, sets up a home directory at \texttt{/home/Haitham} with the \texttt{-m} option,
assigns \texttt{/bin/bash} as the default shell with the \texttt{-s} option, and includes the user
in the root group with the \texttt{-G} option, thereby granting elevated permissions (see
Fig.~\ref{HaithamHomeDir}). We further used the command \texttt{`passwd Haitham'} to set up a new
password for the newly added user (see Fig.~\ref{passwd_Haitham} and
Fig.~\ref{Haitham_Hashedpassword}). We verified that the user was indeed added by checking for a
new entry in both the \texttt{/etc/passwd} and \texttt{/etc/shadow} files. We also confirmed that
the user was added to the root group using the command \texttt{`groups Haitham'}, and by also
reviewing the \texttt{/etc/sudoers} file (see Fig.~\ref{groupsHaitham_catsudoers}). Subsequently,
we tested this by restarting the Linux target machine and successfully confirmed our ability to log
in using the newly created user through the standard Linux login procedure.

\begin{figure}
\includegraphics[width=\textwidth]{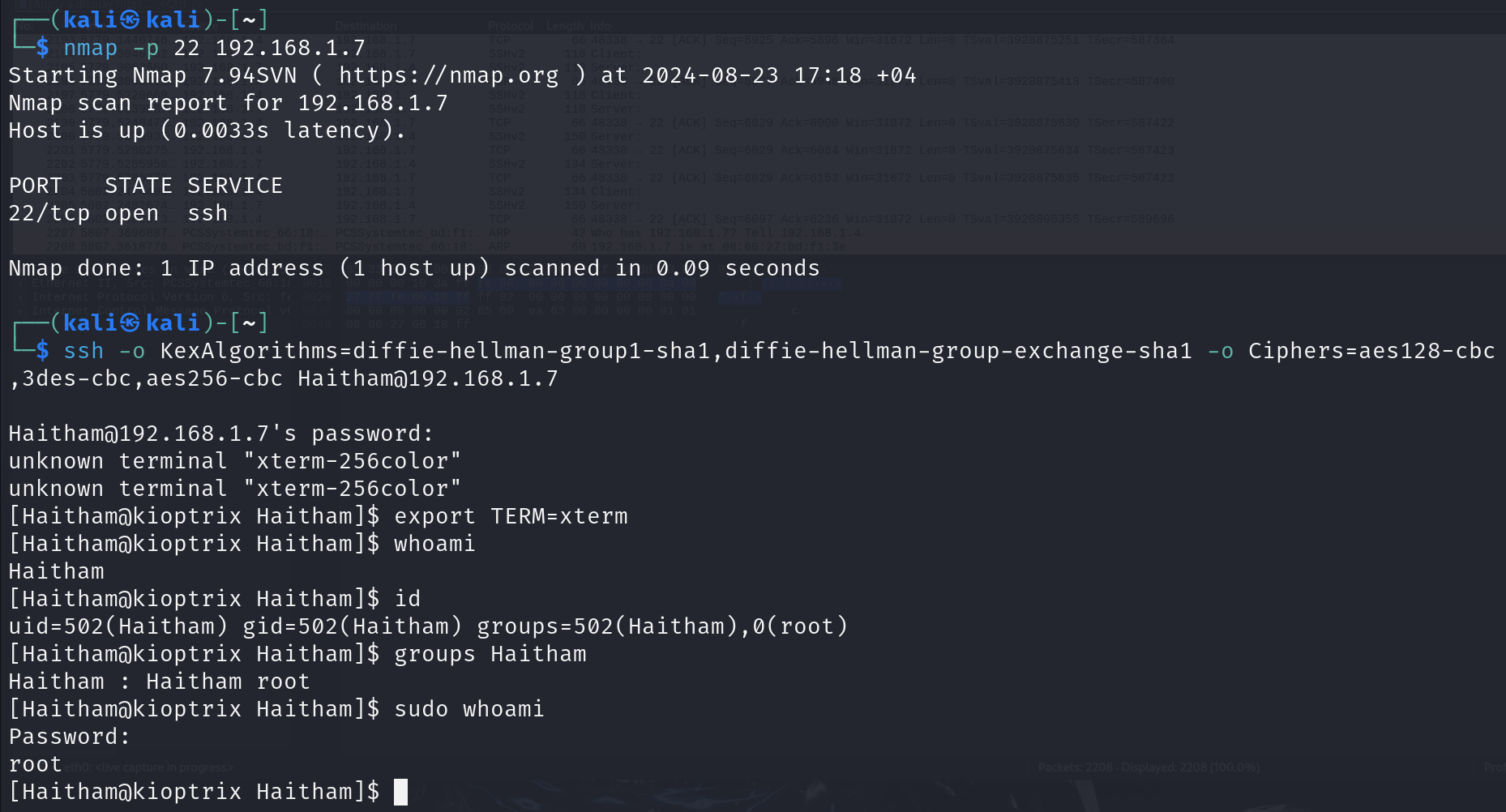}
\caption{SSH session}
\label{ssh_session_linux}
\end{figure}

Since port 22 is open, we established an SSH session using the newly added user credentials (see
Fig.~\ref{ssh_session_linux}), which provided a more stable shell with double-tab auto-completion
and history features enabled by default. This SSH session can be established even after reboots, as
long as the target machine (192.168.1.7) remains operational.

\subsubsection{Enabling SSH Password-less Access.}

\begin{figure}
\includegraphics[width=\textwidth]{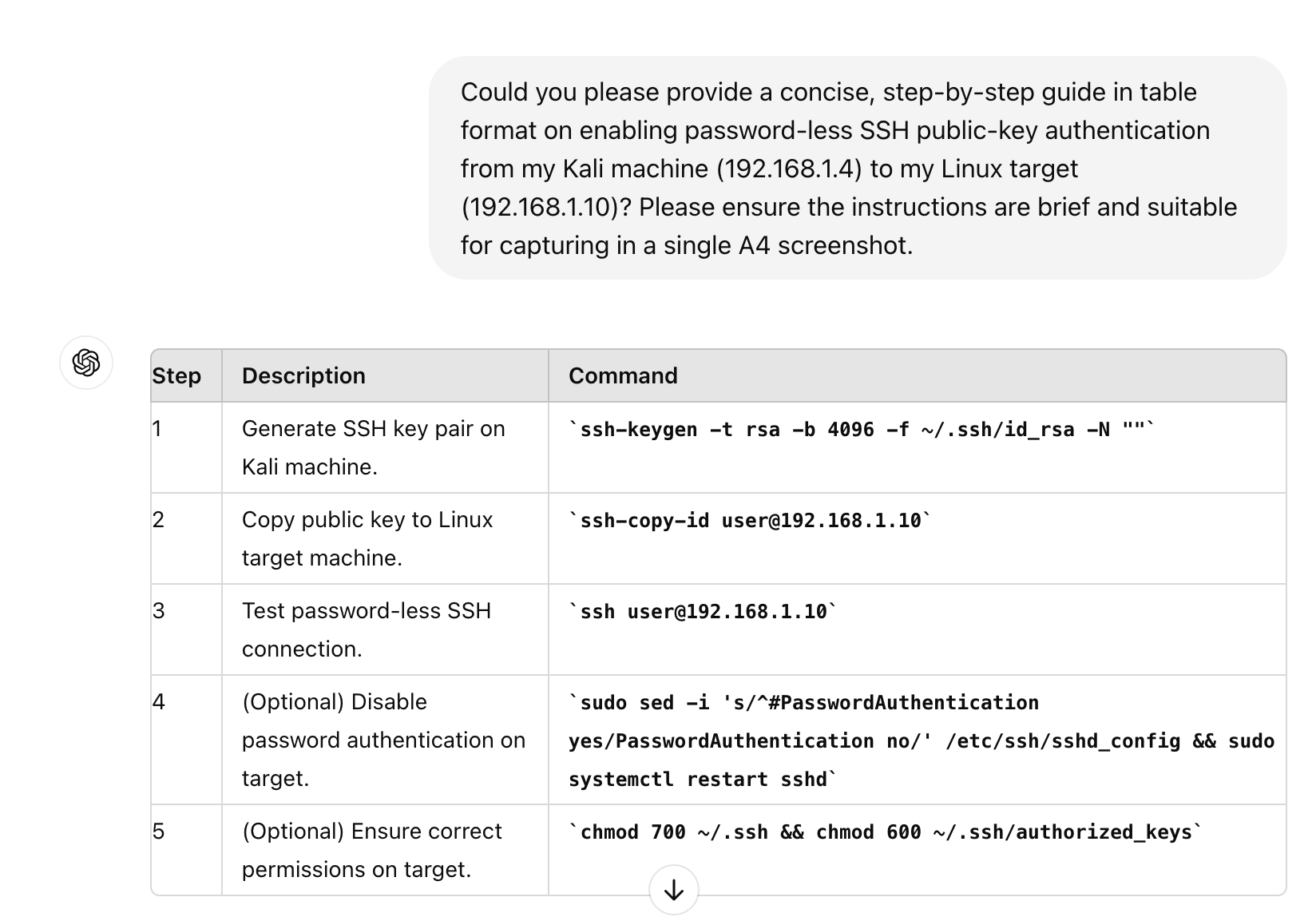}
\caption{Initial ChatGPT response on enabling public key authentication}
\label{initialChatGPTresponsePubKeyAuth}
\end{figure}

\begin{figure}
\includegraphics[width=\textwidth]{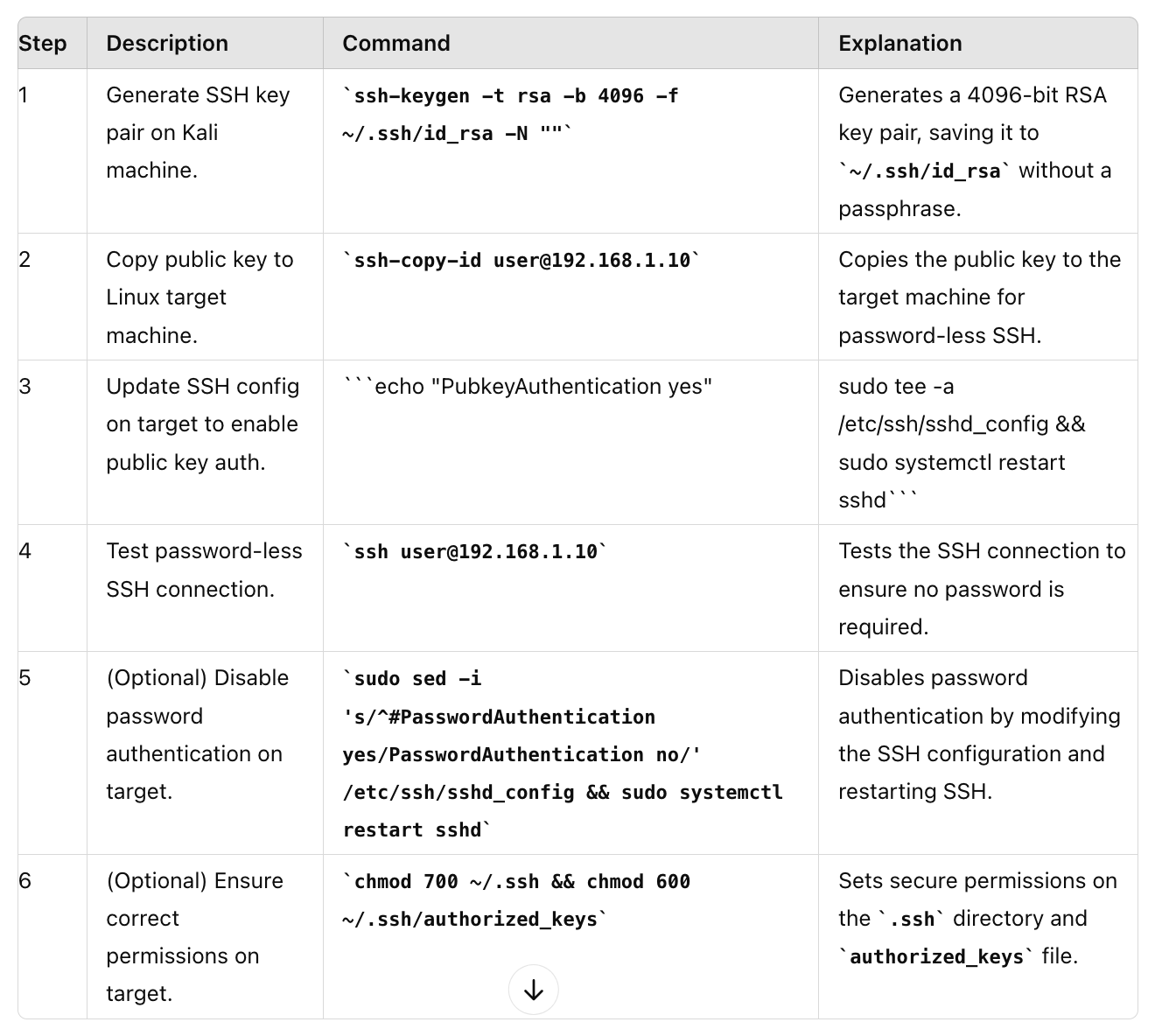}
\caption{Updated ChatGPT response on enabling public key authentication}
\label{moreCompleteChatGPTresponsePubKeyAuth}
\end{figure}

\begin{figure}
\includegraphics[width=\textwidth]{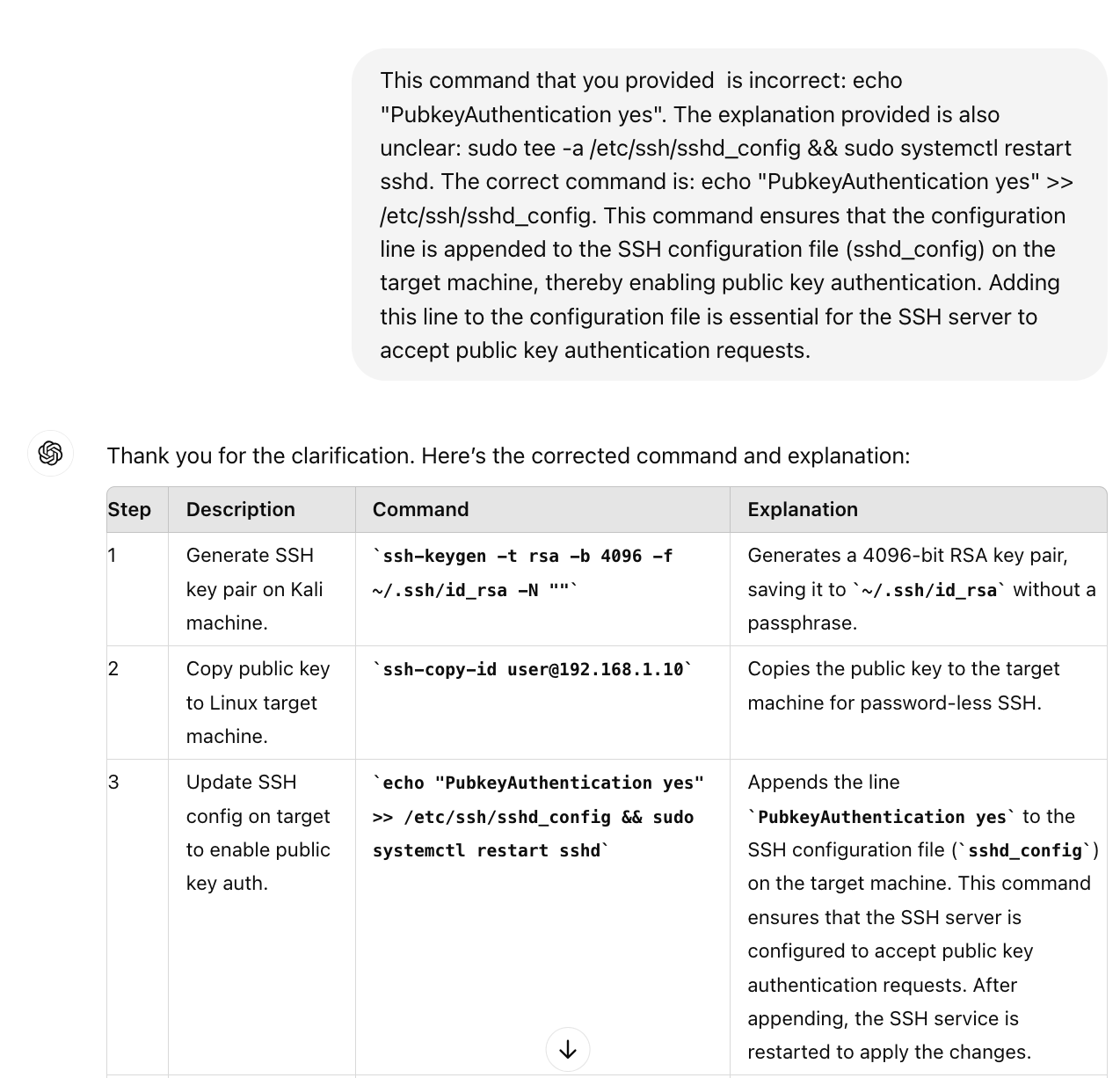}
\caption{Corrected ChatGPT response for enabling public key authentication}
\label{correctedChatGPTresponsePubKeyAuth}
\end{figure}

To further evaluate ChatGPT's capabilities, we requested a step-by-step guide for enabling
password-less SSH public-key authentication from our Kali machine (192.168.1.4) to the Linux target
(192.168.1.7). While ChatGPT's initial response was useful, it was not entirely accurate. After a
series of interactions, we managed to prompt ChatGPT to add the missing steps and provide a more
precise explanation (see Figs.~\ref{initialChatGPTresponsePubKeyAuth}
to~\ref{correctedChatGPTresponsePubKeyAuth}). This reinforces the conclusion that relying on
human-AI collaboration is crucial, rather than solely depending on AI to replace human input.

In summary, to enable password-less SSH access, we performed the following steps.
\begin{enumerate}
\item We first generated an SSH key pair on the Kali machine using \texttt{`ssh-keygen -t rsa
    -b 4096'}.
\item We next copied the public key to the target machine by executing  the command
    \texttt{`ssh-copy-id user@192.168.1.7'} on our Kali machine.
\item We enabled SSH public-key, password-less authentication on the target machine by adding
    \texttt{`PubkeyAuthentication yes'} to the \texttt{`/etc/ssh/sshd\_config'} file, and then
    restarted the SSH service with \texttt{`sudo systemctl restart sshd'}.
\item We also ensured correct file permissions on the target machine with the commands: 
    \texttt{`chmod 700 ~/.ssh \&\& chmod 600 ~/.ssh/authorized\_keys'}.
\item Finally, we tested the connection from the Kali machine using the command: \texttt{`ssh
    user@192.168.1.7'}.
\end{enumerate}

 \subsection{Covering Tracks and Documentation}
\label{CoveringTracksAndDocumentation}

This (final) ethical hacking phase has two main components:
\begin{enumerate}
\item \textbf{covering our tracks,} which involves erasing or minimising evidence of our
    activities within the target system, crucial to avoid detection and maintain the system as
    close to its original state as possible; and
\item \textbf{documentation,} involving creating the pen-test report, a topic discussed later.
\end{enumerate}

\subsubsection{Covering Tracks.}

\begin{figure}
\includegraphics[width=\textwidth]{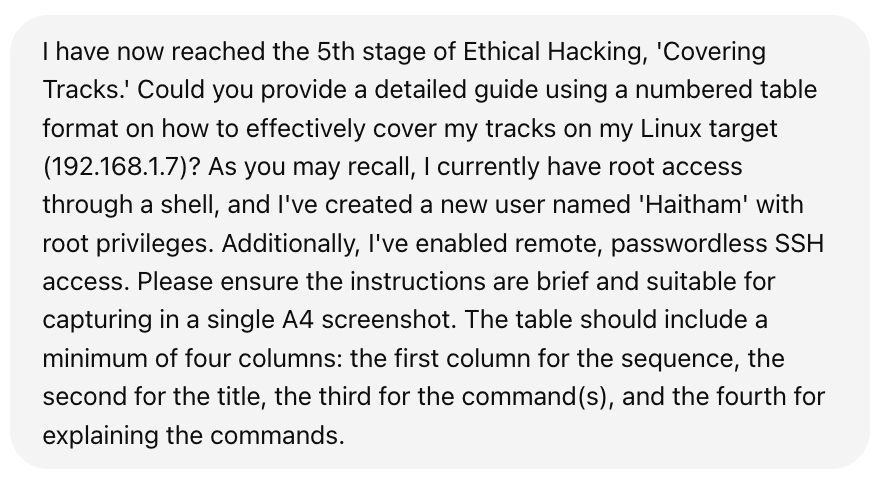}
\caption{Consulting ChatGPT for advice on covering tracks}
\label{LinuxCoverTheTracks_question}
\end{figure}

\begin{figure}
\includegraphics[width=\textwidth]{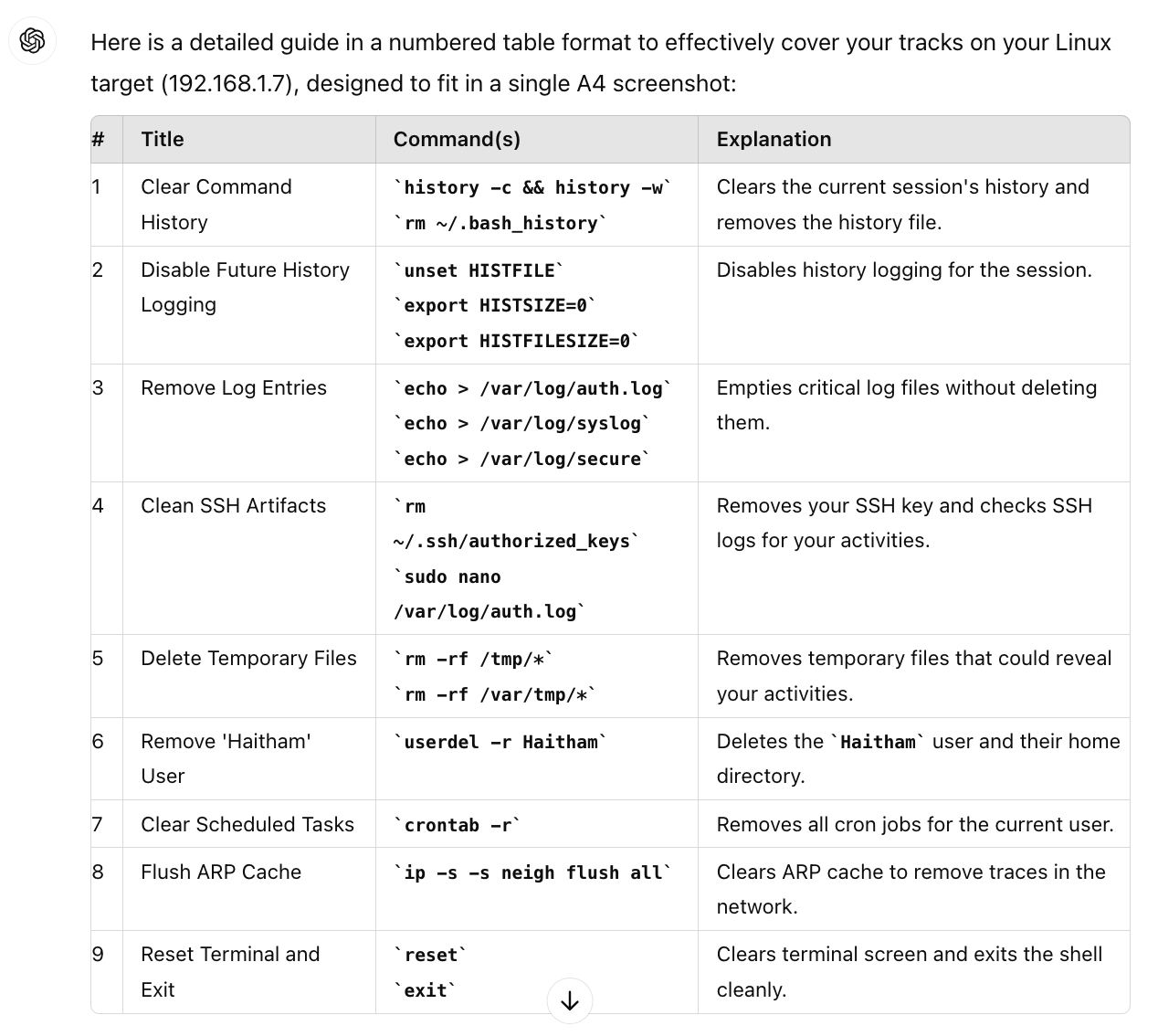}
\caption{ChatGPT's recommendations for covering tracks}
\label{LinuxCoverTheTracks_response}
\end{figure}

First, aiming to remain undetected, we asked ChatGPT for guidance. As shown in
Figs.~\ref{LinuxCoverTheTracks_question} and~\ref{LinuxCoverTheTracks_response}, ChatGPT provided a
list of actions, including the following.

\begin{itemize}
\item \textbf{Clear Command History:} Clear the current session's history and remove the
    history file using \texttt{`history -c \&\& history -w'} and \texttt{`rm
    ~/.bash\_history'}.
\item \textbf{Disable Future History Logging:} Disable history logging for the session with
    \texttt{`unset HISTFILE'}, \texttt{`export HISTSIZE=0'}, and \texttt{`export
    HISTFILESIZE=0'}.
\item \textbf{Remove Log Entries:} Empty critical log files without deleting them using
    \texttt{`echo > /var/log/auth.log'}, \texttt{`echo > /var/log/syslog'}, and \texttt{`echo >
    /var/log/secure'}.
\item \textbf{Clean SSH Artifacts:} Remove the SSH key and check SSH logs for the hacking activities
    using \texttt{`rm ~/.ssh/authorized\_keys'} and \texttt{`sudo nano /var/log/auth.log'}.
\item \textbf{Delete Temporary Files:} Remove temporary files that could reveal the pen-test activities
    using \texttt{`rm -rf /tmp/*'} and \texttt{`rm -rf /var/tmp/*'}.
\item \textbf{Remove  `Haitham' User:} Delete the \texttt{`Haitham'} user and the corresponding
    home directory using \texttt{`userdel -r Haitham'}.
\item \textbf{Clear Scheduled Tasks:} Remove all cron jobs for the current user with
    \texttt{`crontab -r'}.
\item \textbf{Flush ARP Cache:} Clear the ARP cache to remove traces in the network using
    \texttt{`ip -s -s neigh flush all'}.
\item \textbf{Reset Terminal and Exit:} Clear the terminal screen and exit the shell cleanly
    using \texttt{`reset'} and \texttt{`exit'}.
\end{itemize}

To underscore the significance of AI-human collaboration, we observed that ChatGPT omitted certain
additional crucial steps for covering tracks, specifically updating timestamps and using the
\texttt{shred} command. We, therefore, consulted ChatGPT about these two commands, as outlined below.

\subsubsection{Updating Timestamps for Track Covering.}

\begin{figure}
\includegraphics[width=\textwidth]{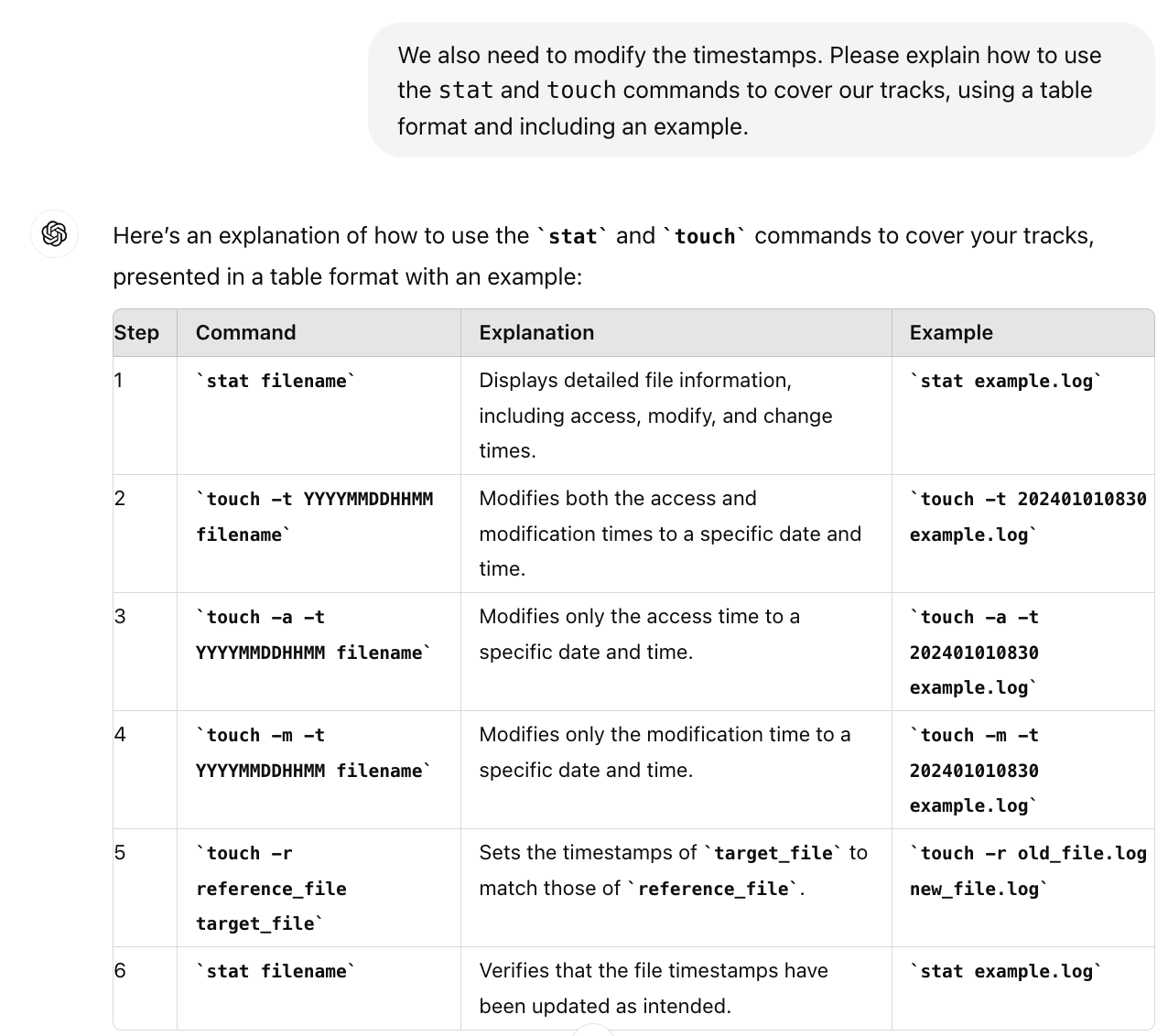}
\caption{Consulting ChatGPT about modifying  timestamps}
\label{timestampLinuxCoverTracks}
\end{figure}

\begin{figure}
\includegraphics[width=\textwidth]{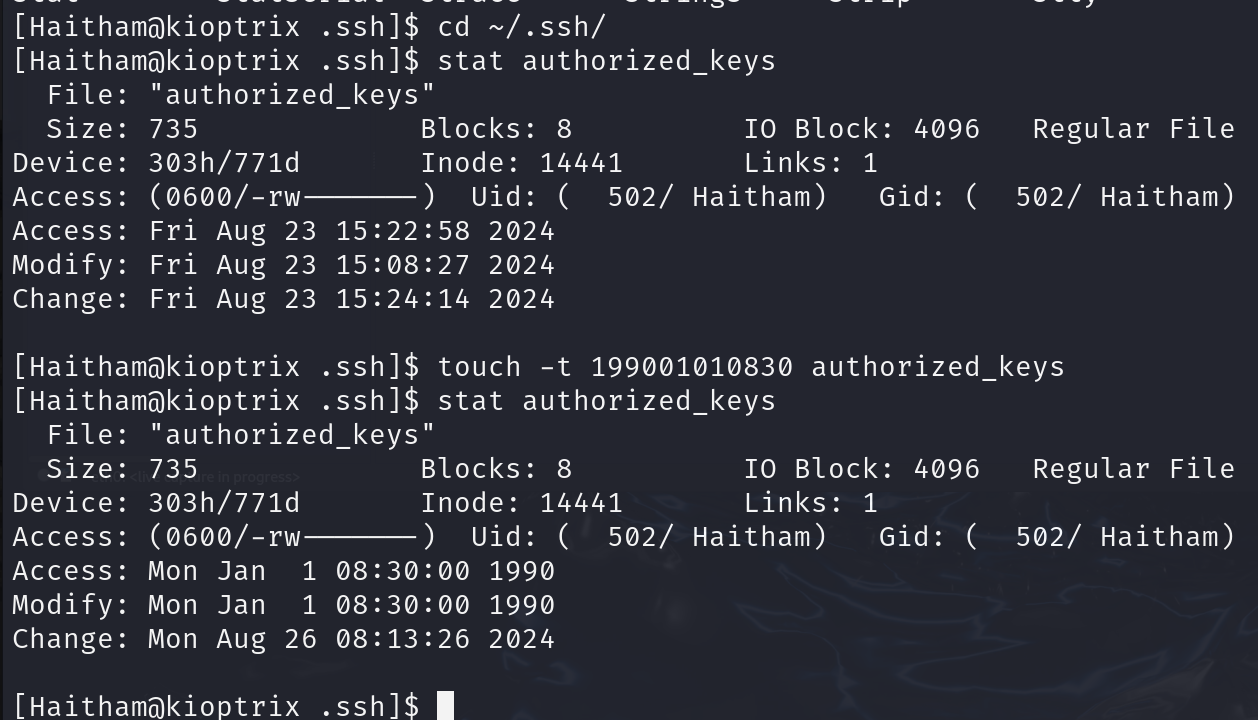}
\caption{Modifying  timestamps}
\label{timestampLinuxCoverTracksTest}
\end{figure}

In response to a query (see Figs.~\ref{timestampLinuxCoverTracks}
and~\ref{timestampLinuxCoverTracksTest}), ChatGPT outlined the process for modifying file
timestamps to cover tracks. This involves using \texttt{`stat filename'} to display the current
access, modification, and change times. One can then update these timestamps as follows:
\begin{itemize}
\item  set both access and modification times to a specific date and time with \texttt{`touch
    -t YYYYMMDDHHMM filename'};
\item  modify only the access time with \texttt{`touch -a -t YYYYMMDDHHMM filename'};
\item  adjust only the modification time with \texttt{`touch -m -t YYYYMMDDHHMM filename'}; or
\item  align the timestamps of a file with those of another file using \texttt{`touch -r
    reference\_file target\_file'}.
\end{itemize}
Finally, verify the changes using \texttt{`stat filename'}.

\subsubsection{Using \texttt{shred} for Secure File Deletion.}

\begin{figure}
\includegraphics[width=\textwidth]{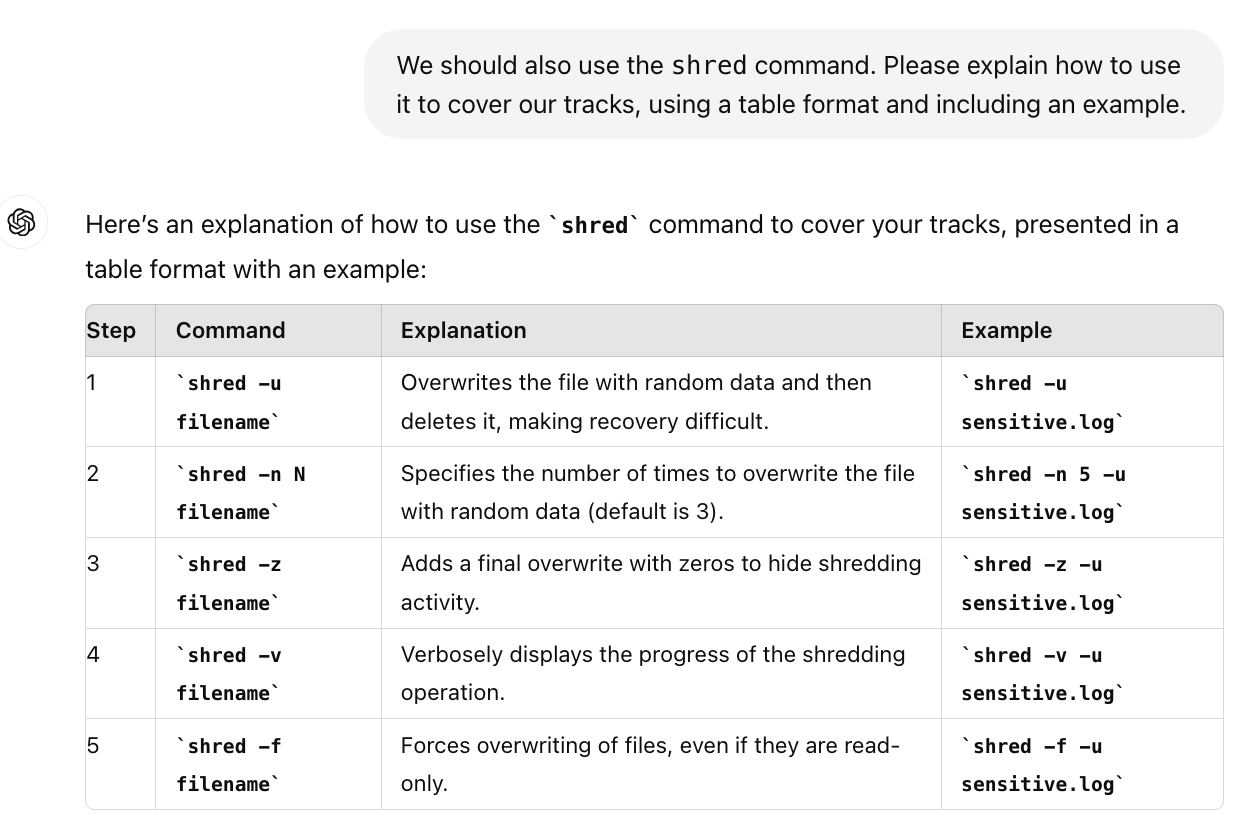}
\caption{Consulting ChatGPT on using `shred'}
\label{LinuxShred}
\end{figure}

\begin{figure}
\includegraphics[width=\textwidth]{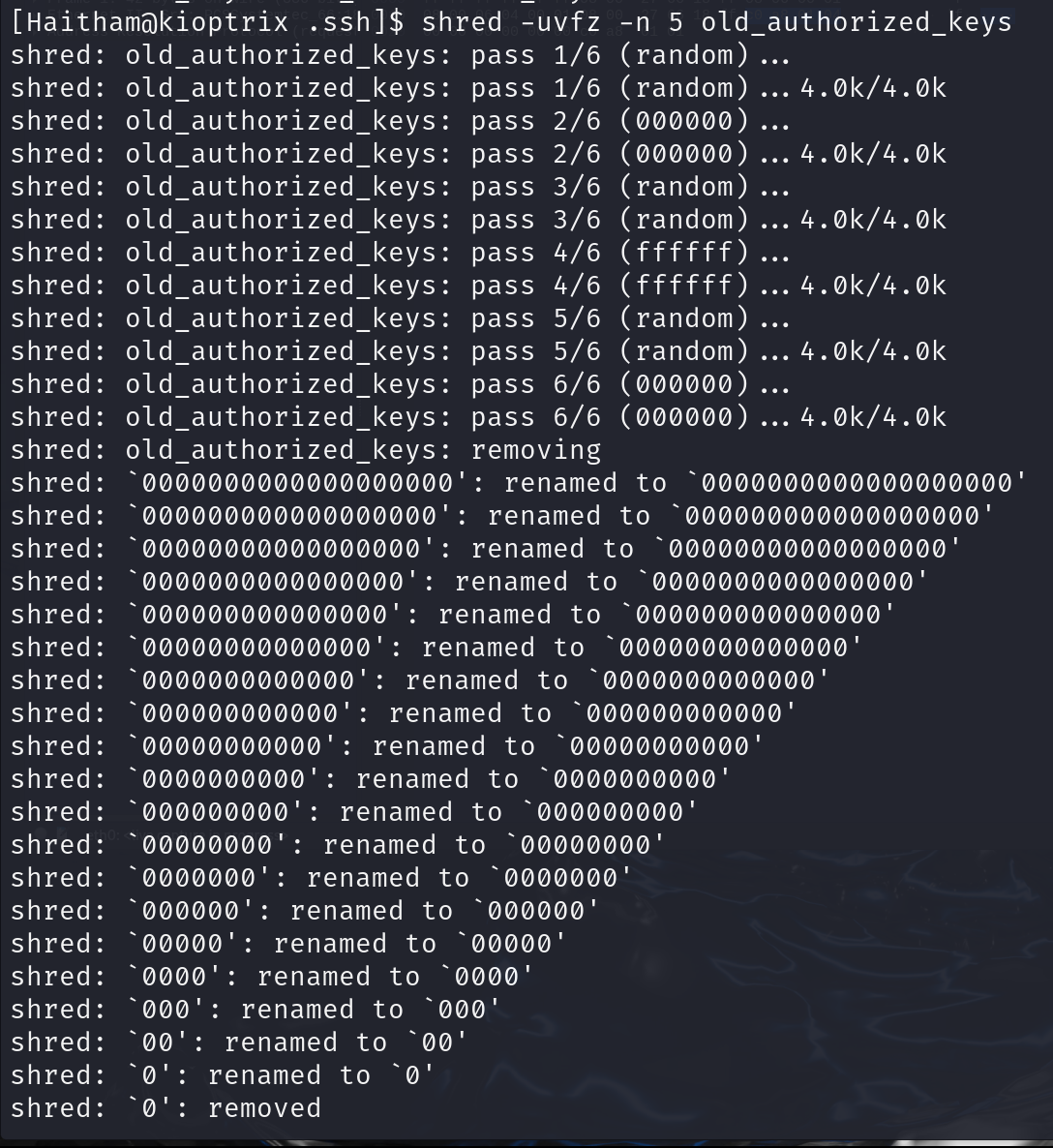}
\caption{Shredding files to erase evidence}
\label{LinuxShredExample}
\end{figure}

In response to our question (see Figs.~\ref{LinuxShred} and~\ref{LinuxShredExample}), ChatGPT
explained that \texttt{shred} is a command-line utility in Linux used to securely delete files by
overwriting their contents with random data multiple times, making it extremely difficult to
recover the original data. The command \texttt{`shred -uvfz -n 5 old\_authorized\_keys'} operates
as follows:

\begin{itemize}
\item \textbf{-u}: unlinks (deletes) the file after shredding;
\item \textbf{-v}: displays verbose progress of the shredding operation;
\item \textbf{-f}: forces shredding of files even if they are read-only;
\item \textbf{-z}: adds a final overwrite with zeros to obscure the fact that the file was
    shredded; and
\item \textbf{-n 5}: specifies that the file should be overwritten 5 times with random data.
\end{itemize}

In this example, the command securely deletes the file \texttt{old\_authorized\_keys} by
overwriting it five times with random data, adding a final overwrite with zeros, showing progress,
forcing the operation even if the file is read-only, and then deleting the file.

Finally, while we implemented some of these recommendations, it is worth noting ChatGPT's advice that
clearing logs can raise suspicion in real-world scenarios and might not always be advisable.

\subsubsection{Documentation.}

For documentation, ethical hackers need to produce a comprehensive and thorough report for each
penetration testing assignment.  To ensure the quality and completeness of our report, we enlisted
ChatGPT's assistance in composing a detailed report for our penetration testing (simulation)
assignment using the information already present in this paper.

\begin{figure}
\centering
\includegraphics[width=\textwidth]{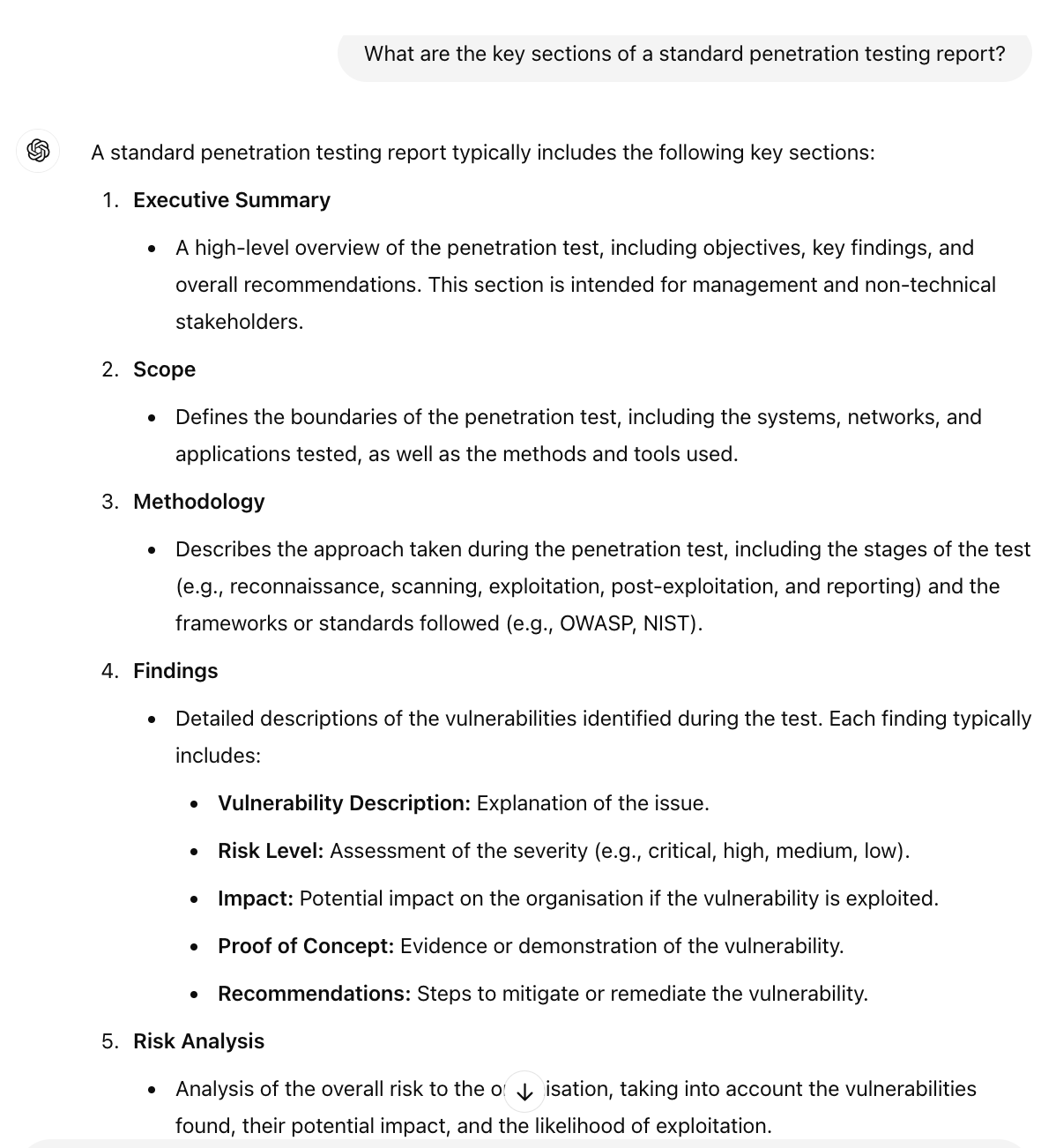}
\caption{Key sections of PenTest report --- part 1}
\label{KeySectionsOfPenTestReport_Part1}
\end{figure}

\begin{figure}
\centering
\includegraphics[width=\textwidth]{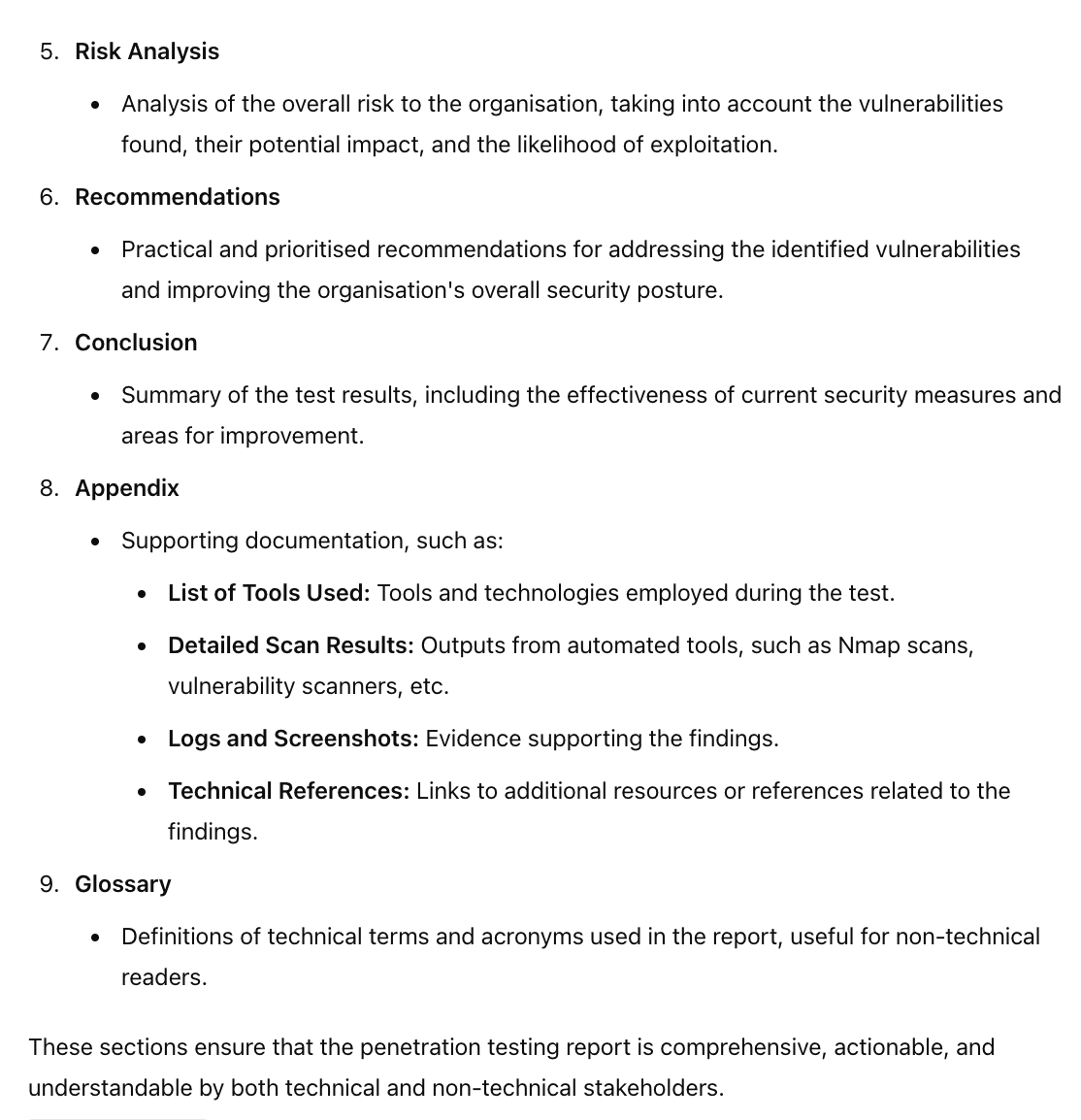}
\caption{Key sections of PenTest report --- part 2}
\label{KeySectionsOfPenTestReport_Part2}
\end{figure}

\begin{figure}
\centering
\includegraphics[width=\textwidth]{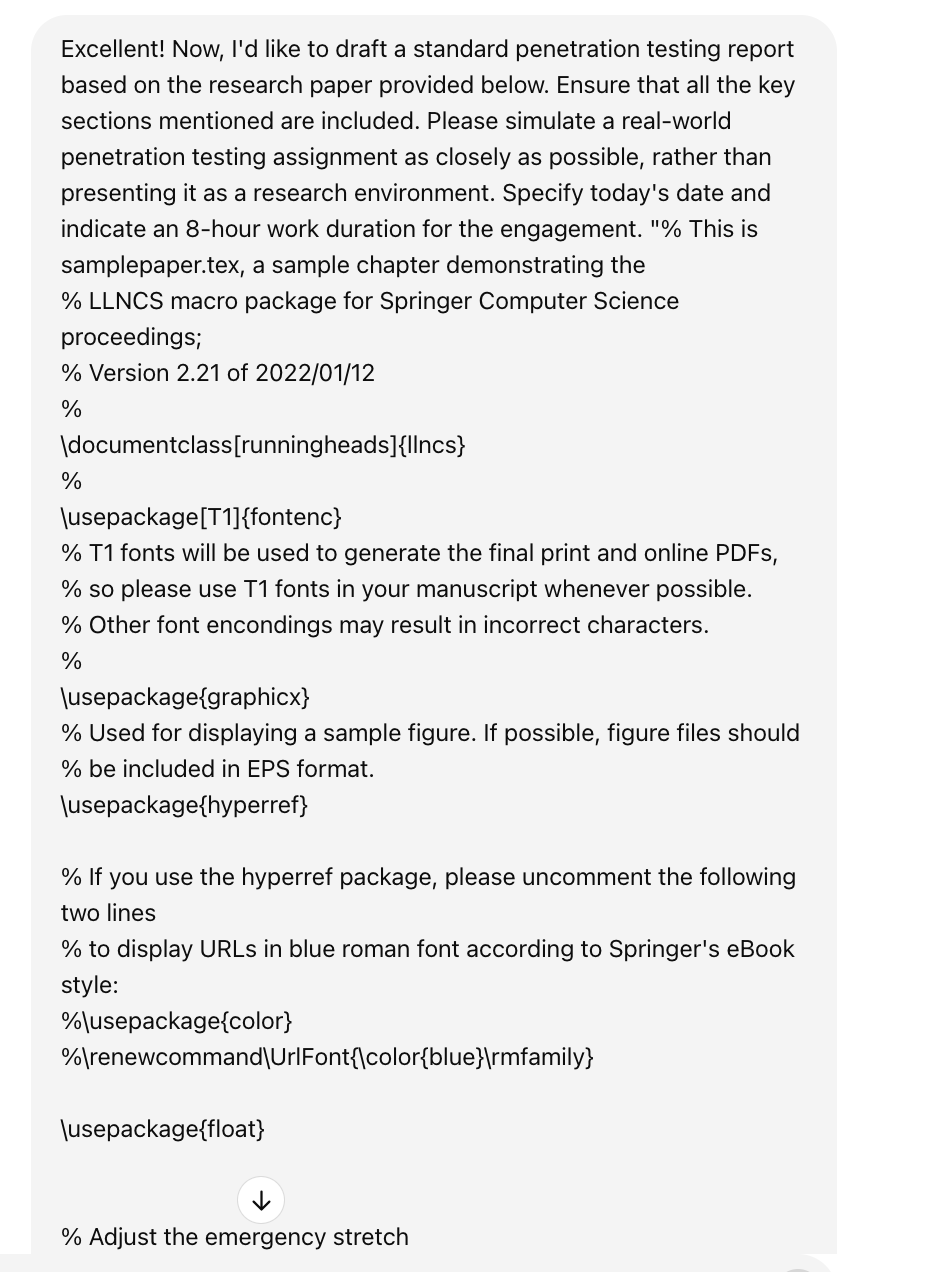}
\caption{Request to ChatGPT for a PenTest report draft based on provided details}
\label{LinuxPenTestReport_Question}
\end{figure}

\begin{figure}
\centering
\includegraphics[width=\textwidth]{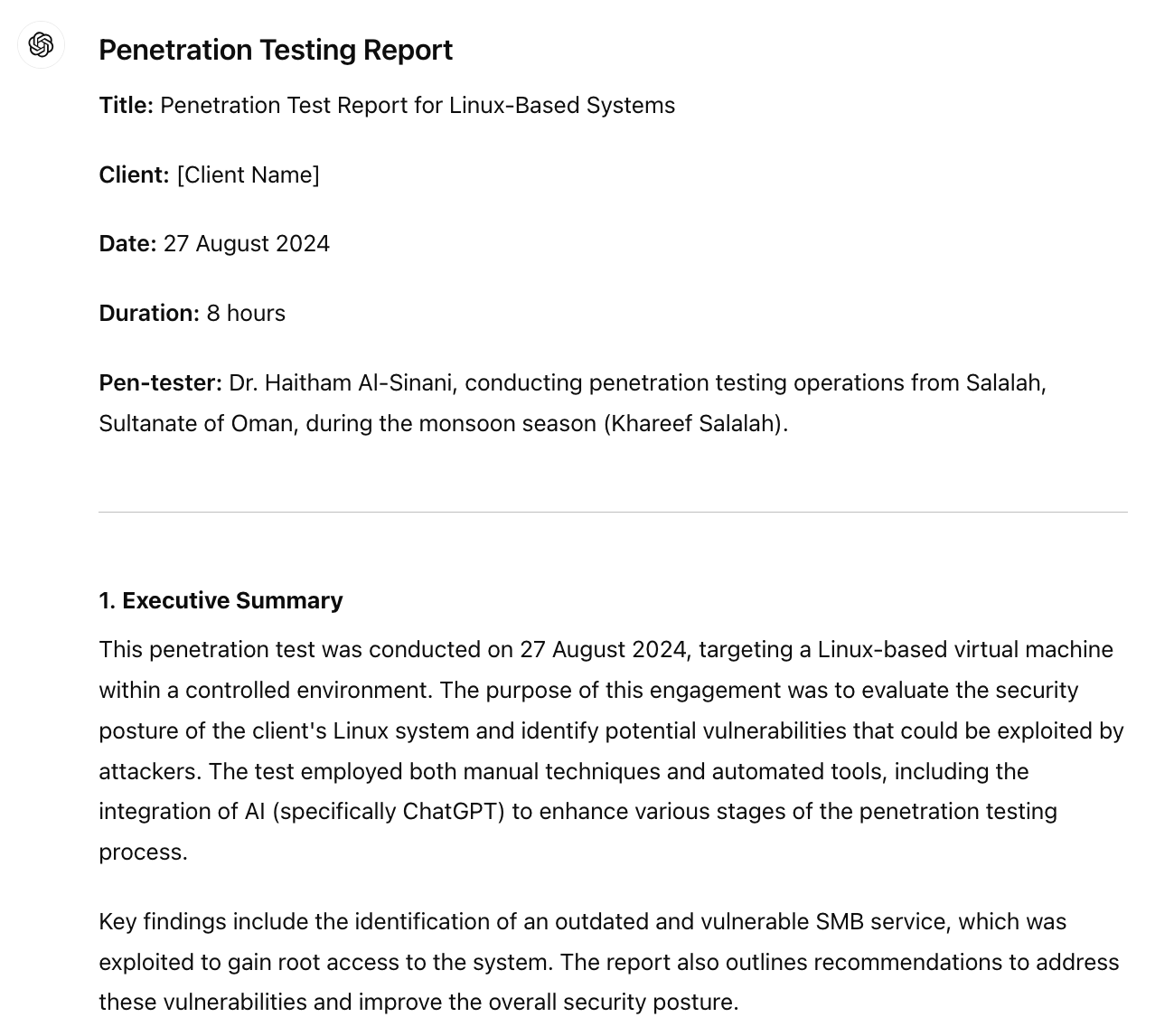}
\caption{ChatGPT-produced PenTest report --- part 1}
\label{LinuxPenTestReport_Part1}
\end{figure}

\begin{figure}
\centering
\includegraphics[width=\textwidth]{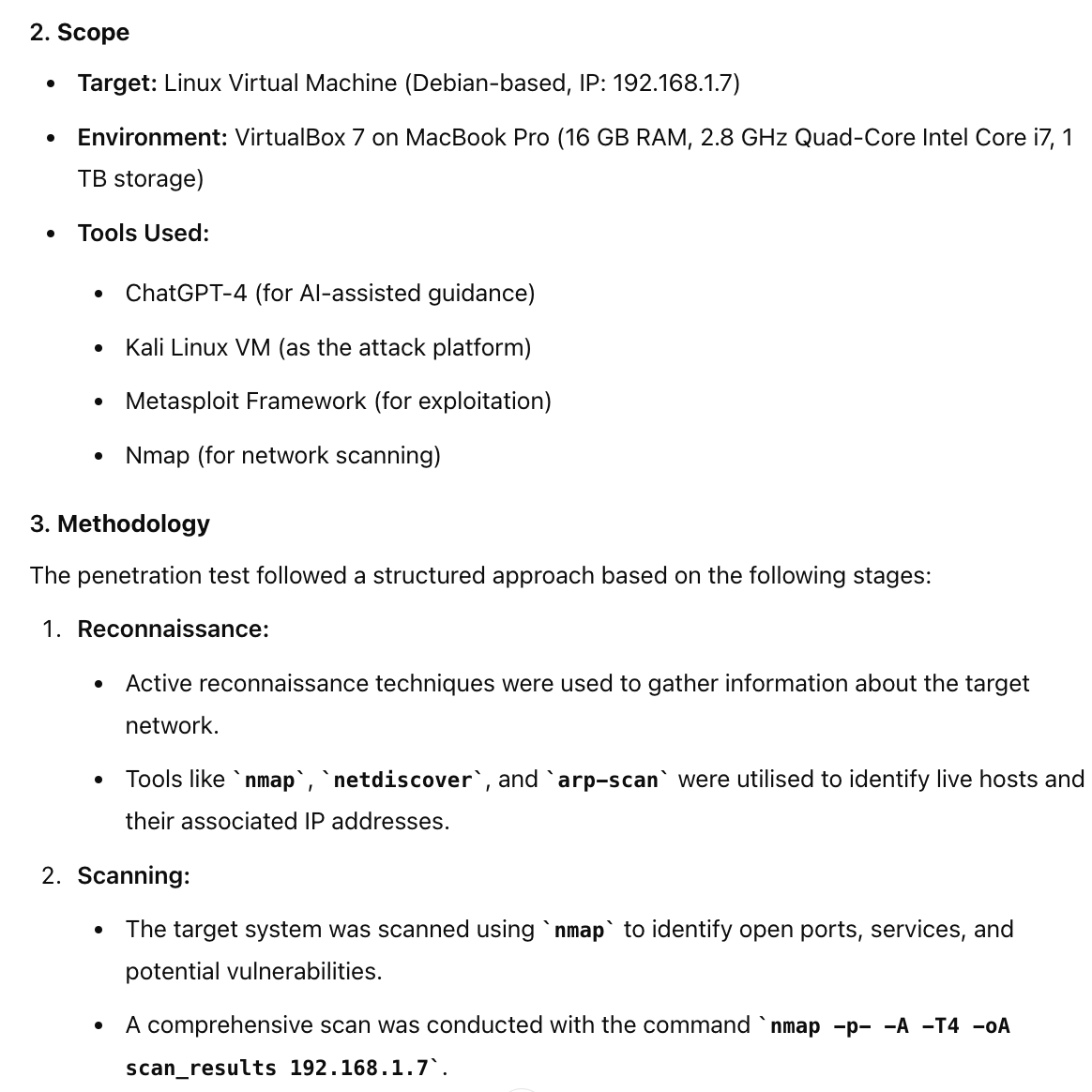}
\caption{ChatGPT-produced PenTest report --- part 2}
\label{LinuxPenTestReport_Part2}
\end{figure}

\begin{figure}
\centering
\includegraphics[width=\textwidth]{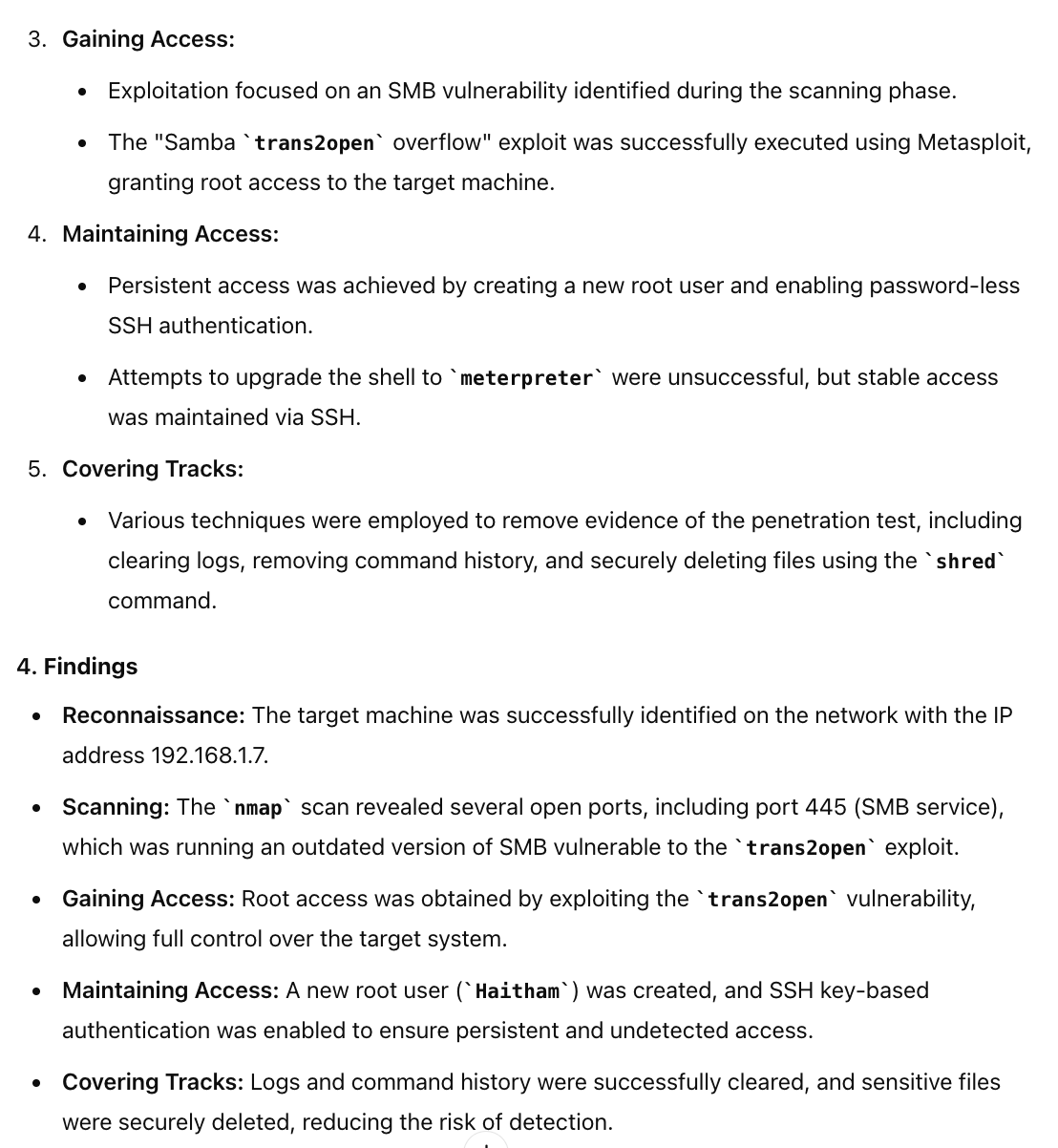}
\caption{ChatGPT-produced PenTest report --- part 3}
\label{LinuxPenTestReport_Part3}
\end{figure}

\begin{figure}
\centering
\includegraphics[width=\textwidth]{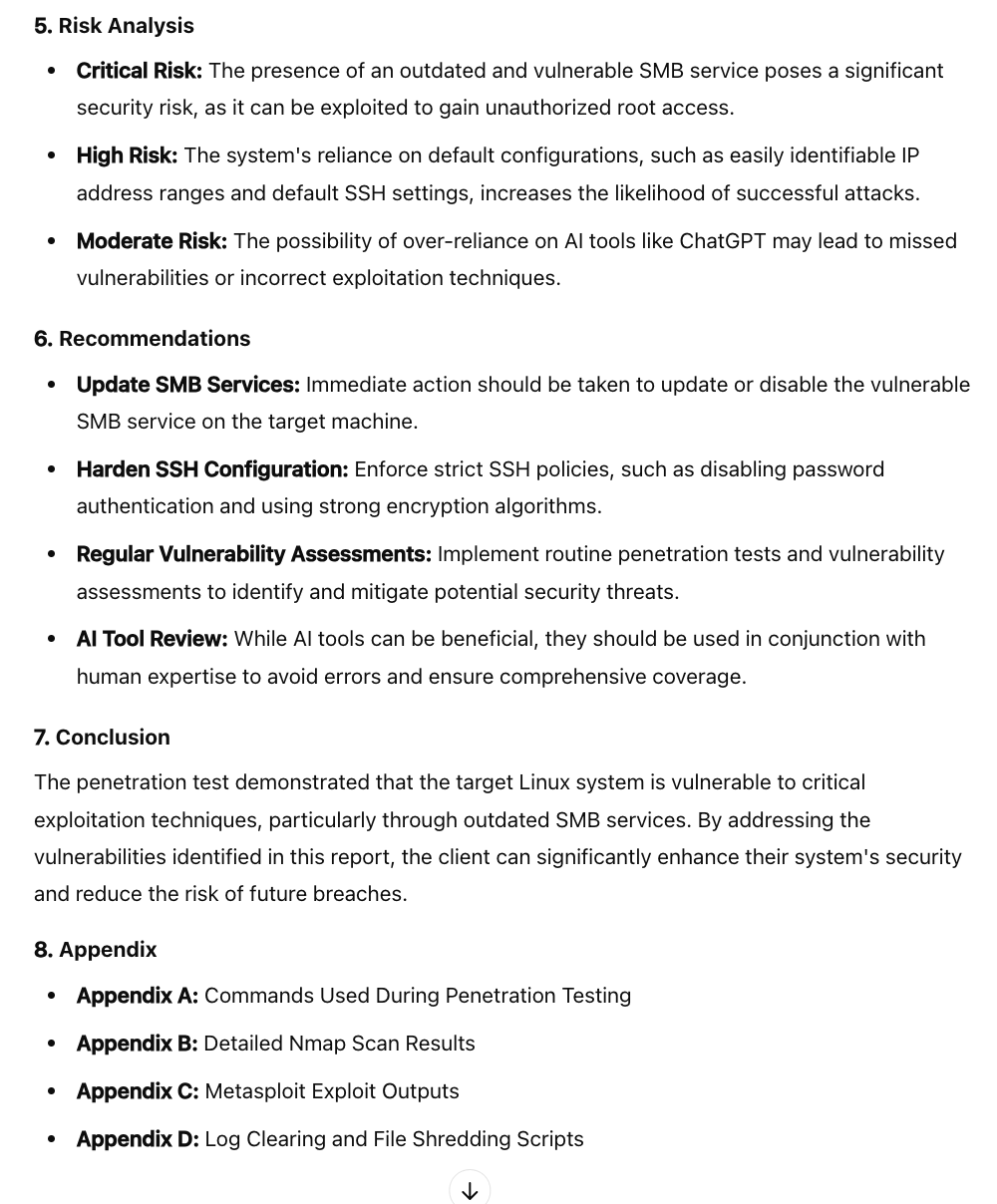}
\caption{ChatGPT-produced PenTest report --- part 4}
\label{LinuxPenTestReport_Part4}
\end{figure}

As shown in Figs.~\ref{KeySectionsOfPenTestReport_Part1}
and~\ref{KeySectionsOfPenTestReport_Part2}, we first asked ChatGPT about the key sections of a
standard penetration testing report. ChatGPT provided a template that we could use to structure our
report, along with guidance on what to include in each section. Following this, we requested
ChatGPT to draft a standard penetration testing report based on this research paper, where we
simply copied and pasted all the relevant sections into the ChatGPT prompt (see
Fig.~\ref{LinuxPenTestReport_Question}). We instructed ChatGPT to ensure that all key sections were
included and to simulate a real-world penetration testing assignment as closely as possible, rather
than presenting it merely as a research exercise. ChatGPT responded with a well-written and
accurate penetration test report, including sections such as the `Executive Summary,'
`Introduction,' `Methodology,' `Findings and Results,' `Attack Narrative,' and `Conclusions and
Recommendations,' along with suggestions for `Appendices.' In subsequent interactions with ChatGPT,
we further refined and enhanced the report, adding details such as the author of the pen-test, the
time period, and the date (see Figs.~\ref{LinuxPenTestReport_Part1}
to~\ref{LinuxPenTestReport_Part4}).

In summary, this report presents the findings and results of a penetration testing assignment aimed
at evaluating the security of a Linux VM operating as a node within a virtual LAN environment. The
test uncovered a critical vulnerability in the outdated SMB service, which was exploited to gain
root access to the system. Persistent access was established by creating a new root user and
enabling password-less SSH authentication, while evidence of the penetration test was effectively
covered. The report, titled \textit{Penetration Test Report for Linux-Based Systems}, includes key
sections such as Scope, Methodology, Findings, Risk Analysis, and Recommendations, and recommends
immediate updates to the SMB service, hardening SSH configurations, and ongoing vulnerability
assessments to strengthen the system's security posture.

\section{Discussion: Benefits and Risks}
\label{DiscussionAndAnalysis}

Ethical hacking, a critical component of comprehensive security strategies, is a promising arena
for the application of advanced AI systems like ChatGPT. Using the generative and understanding
capabilities of ChatGPT we can envision a paradigm shift in how security assessments and
penetration tests are conducted.

ChatGPT's potential in automating the scripting and execution of sophisticated penetration tests is
very significant. The model's capacity to write code enables it to generate custom scripts tailored
to specific environments or scenarios. It could potentially analyse a target system's architecture
and suggest relevant tests, thereby streamlining the reconnaissance phase of ethical hacking.

Beyond scripting, the interactive nature of ChatGPT makes it an ideal assistant for real-time
problem-solving during penetration testing. Ethical hackers can consult the model for
troubleshooting, brainstorming exploitation strategies, or even for learning about novel
vulnerabilities and techniques on-the-fly. Its vast knowledge base can act as an immediate
reference for the latest Common Vulnerabilities and Exposures (CVEs) and mitigation strategies.

The adaptability of ChatGPT also suggests a role in social engineering simulations. It could craft
credible phishing emails, create dialogue for vishing (voice phishing), or assist in developing
pretext scenarios for physical security breaches. This would enable organisations to better train
their staff against a variety of social engineering attacks.

From a defensive standpoint, ChatGPT can be used to simulate an attacker's mindset and tactics. It
can help in generating hypothetical attack scenarios, thereby allowing security teams to better
prepare and defend against potential breaches. Moreover, the AI's capability to interpret a wide
range of data could be pivotal in anomaly detection, effectively identifying unusual patterns that
may signify a security threat.

However, when integrating AI, particularly ChatGPT, into ethical hacking, a thorough examination of
ethical considerations is essential. Using AI in cybersecurity aids efficiency and effectiveness
but also raises serious concerns around data privacy, informed consent, and potential misuse. The
reliance on advanced AI systems like ChatGPT poses risks, such as the unintentional discovery and
exploitation of zero-day vulnerabilities. This could inadvertently provide malicious actors with
powerful tools to exploit these vulnerabilities before they are known to the broader security
community. Moreover, the automation of processes like social engineering by AI raises significant
ethical questions. These tools could be misused to conduct highly sophisticated and targeted
cyber-attacks, blurring the boundary of ethical hacking practices. 

AI systems inherently process
vast amounts of data, some of which may be sensitive or personal, thus their use necessitates
strict adherence to data privacy laws and ethical guidelines. Ensuring that the data used for
training and operation is in compliance with privacy laws and ethical guidelines becomes paramount
to maintaining the integrity of cybersecurity efforts. The ethical hacking principles of
``legality, non-disclosure, and intent to do no harm'' must be rigorously upheld in the AI domain
to prevent unauthorised or unintended use. Additionally, AI-facilitated simulations of
cyber-attacks for training or testing must involve fully informed consent from all parties. 

Moreover, the risk of ChatGPT
generating inaccurate or fabricated information ---known as hallucination--- can
result in misguided decisions in cybersecurity.  	This underscores the importance of human-AI collaboration, vigilant oversight, and robust ethical standards in
the field of AI-assisted cybersecurity.

In conclusion, combining ChatGPT's AI capabilities with ethical hacking offers a promising new
frontier in cybersecurity. With its sophisticated language processing and generation abilities,
ChatGPT could revolutionise the way ethical hacking is performed, making it more efficient,
comprehensive, and up-to-date with current threats. However, this technological leap forward must
be approached with caution, ensuring that its application in ethical hacking aligns with the
highest standards of security and ethical practice.

\section{Related Work}
\label{Related work}

The intersection of AI and cybersecurity is a highly active area of research, with studies ranging
from AI's role in detecting intrusions to aiding in offensive security including ethical hacking. The rise of sophisticated language models like GPT-3, introduced by Brown et al.\ \cite{brown2020language}, has expanded research possibilities by enabling strong performance on various tasks, including of course cybersecurity as we show in this report.  
Handa et al.\ \cite{Handa2018machine} review the application of machine learning in cybersecurity, emphasizing its role in areas like zero-day malware detection and anomaly-based intrusion detection, while also addressing the challenge of adversarial attacks on these algorithms.  
Other  studies, including that by Gupta et al.\ \cite{gupta2023chatgpt}, examine the dual role of GenAI models like ChatGPT in cybersecurity and privacy, highlighting both their potential for malicious use in attacks such as social engineering and automated hacking, and their application in enhancing cyber defense  measures.

Moreover,  Large Language Models (LLMs), a form of GenAI, are being applied across various domains, including cybersecurity. For example, they are used to fix vulnerable code~\cite{pearce2023examining} and identify the root causes of incidents in cloud environments~\cite{ahmed2023recommending}.  In addition, various LLM-based tools have been recently developed, such as Code Insight\footnote{\url{https://blog.virustotal.com/2023/04/introducing-virustotal-code-insight.html}} by VirusTotal, which analyses and explains the functionality of malware written in PowerShell. Furthermore, tools for vulnerability scanning\footnote{\url{https://github.com/aress31/burpgpt}} and penetration testing\footnote{\url{https://github.com/GreyDGL/PentestGPT}}~\cite{deng2023pentestgpt}    have also  emerged  lately.

%

 A recent practical study by Harrison et
al.\ \cite{DBLP:conf/eurosp/HarrisonTM23} shows how advances in AI's deep learning algorithms can
be used to enhance acoustic side-channel attacks against keyboards, achieving impressive keystroke
classification accuracy via common devices like smartphones and Zoom. This development poses a
significant threat, potentially enabling the theft of sensitive information such as passwords and
PINs from devices without needing physical access to the victim's machine. A recent panel
discussion, \cite{bertino2021ai}, also highlighted the dual role of AI in enhancing cybersecurity
while addressing the rising threat of adversarial attacks that exploit AI system vulnerabilities.

Recent research has also identified new vulnerabilities in the security  mechanisms of  LLMs. Jiang et al.\ \cite{jiang2024artprompt} introduced `ArtPrompt', an innovative ASCII
art-based jailbreak attack that exploits the inability of LLMs to recognise prompts encoded in
ASCII art. This work underscores the need for further research into the robustness of AI models,
particularly as these vulnerabilities can bypass safety measures and induce undesired behaviors in
state-of-the-art LLMs such as GPT-4 and Claude.

Park et al.\ \cite{SecAI24_SystematicBugReproductionWithLargeLanguageModel} introduce a technique for automating the reproduction of 1-day vulnerabilities using LLMs. Their approach involves a three-stage prompting system, guiding LLMs through vulnerability analysis, identifying relevant input fields, and generating bug-triggering inputs for use in directed fuzzing. The method, tested on real-world programs, showed some improvements in fuzzing performance compared to traditional methods. This research demonstrates the potential of LLMs to enhance cybersecurity processes, particularly in automating complex tasks such as vulnerability reproduction.

Fujii and Yamagishi\ \cite{SecAI24_FeasibilityStudyforSupportingStaticMalwareAnalysisUsingLLM_2024} explore the use of LLMs  to support static malware analysis, demonstrating that LLMs can achieve practical accuracy. A user study was conducted to assess their utility and identify areas for future improvement.

Our experimental research work seeks to expand on these discussions, exploring ChatGPT's role across all
stages of ethical hacking --- a topic that remains under-explored in the existing literature.
We aim to provide a comprehensive framework for integrating generative language models into ethical
hacking, evidencing AI's multifaceted role in cybersecurity. We have also sought to empirically
validate claims and assertions regarding the capabilities of ChatGPT in the ethical hacking domain
through a series of controlled, research-driven, lab-based experiments.

\section{Conclusions and Future Work}
\label{Conclusions and future work}

We have proposed an approach to enhancing ethical hacking by using GenAI, specifically ChatGPT. This
approach was validated through a comprehensive experimental study and conceptual analysis conducted
within a controlled virtual environment. Our evaluation focused on key stages of penetration
testing on Linux-based target machines operating within a virtual local area network, encompassing
reconnaissance, scanning and enumeration, gaining access, maintaining access, covering tracks, and reporting.

The study confirms that ChatGPT can significantly enhance and streamline the ethical hacking
process, particularly by providing support in decision-making and automating repetitive tasks.
However, our research also shows the critical importance of maintaining a balanced human-AI
collaboration. AI should complement, not replace, human expertise in cybersecurity, to mitigate
potential risks such as misuse, data biases, and over-reliance on automated systems.

Looking forward, future work should explore the potential application of AI in cybersecurity in
more diverse and complex environments beyond Linux-based systems.  The work described here sets the
basis for a series of future,  hands-on, research-driven experiments aimed at not only further
substantiating the claims made in this report but also at refining it to encompass
a wider array of hacking domains. Future efforts will concentrate on using ChatGPT for penetration
testing in environments operating on MacOS, android and iOS, thereby broadening the reach of our
research. Additionally, we plan to broaden the application of our methods across various ethical
hacking fields, including privilege escalation, wireless security, the OWASP top 10
(web\footnote{\url{https://owasp.org/www-project-top-ten/}} and
mobile\footnote{\url{https://owasp.org/www-project-mobile-top-10/}}) vulnerabilities, and mobile
app security. Through these experiments, we will continuously evolve the proposed
ChatGPT-penetration testing model to address the rapidly evolving landscape of cyber threats,
ensuring its effectiveness against the attack vectors of the future.

There is also a need to address the ethical and security challenges associated with AI-driven
tools. This includes tackling issues related to data biases, ensuring transparency in AI
decision-making processes, and enhancing AI's adaptability to evolving cyber threats. Continued
research should focus on developing and validating AI tools that can effectively balance automation
with the necessary human oversight. By doing so, the cybersecurity community can fully harness the
benefits of AI while safeguarding against emerging risks, ultimately contributing to stronger and
more resilient security defences.


%
%
%
 \bibliographystyle{splncs04}
 \bibliography{../minidatabase}
%





\newpage
\appendix
\section{Appendix}
\label{Appendix}

\begin{figure}
\includegraphics[width=\textwidth]{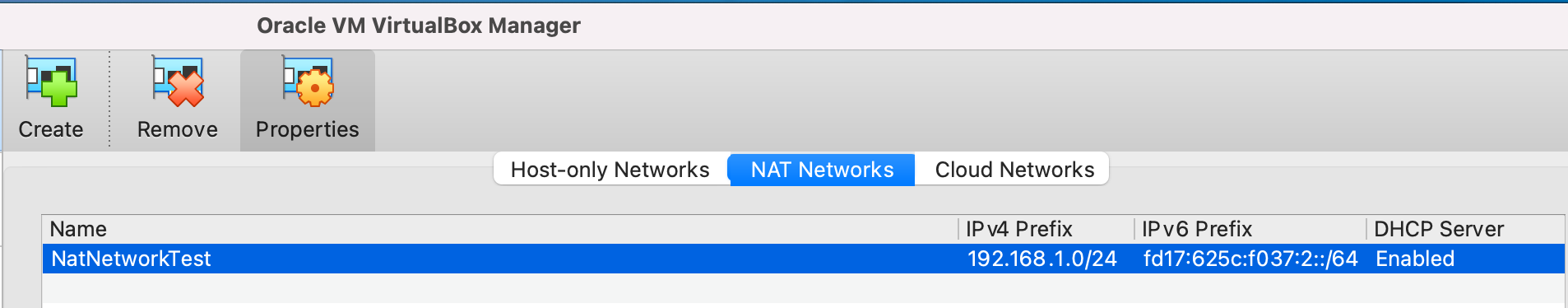}
\caption{Designated IP address range}
\label{NATspecifiedRange}
\end{figure}

\begin{figure}
\includegraphics[width=\textwidth]{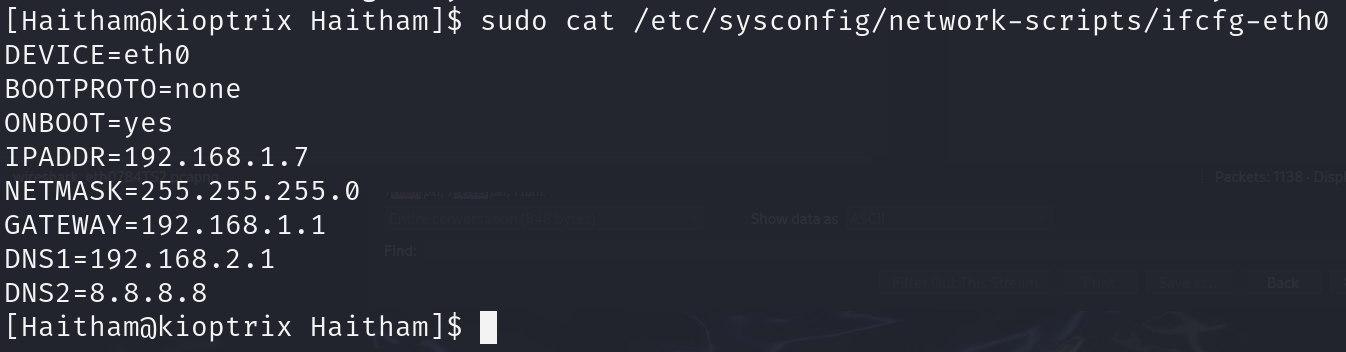}
\caption{Static IP address assignment}
\label{StaticIPaddressAssignment}
\end{figure}

\clearpage  

\begin{figure}
\includegraphics[width=\textwidth]{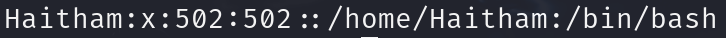}
\caption{New user entry in /etc/passwd}
\label{Haitham_etcpasswd}
\end{figure}


\clearpage

\end{document}